\documentclass{statsoc}

\usepackage[a4paper]{geometry}
\usepackage{graphicx}
\usepackage[textwidth=8em,textsize=small]{todonotes}
\usepackage{natbib}

\usepackage{amssymb, amsmath}
\usepackage{bm}
\usepackage{tikz}

\usepackage[flushleft]{threeparttable}
\usepackage{threeparttable}

\title[Coevolving Latent Space Network with Attractors Model for Polarization]{Disentangling positive and negative partisanship in social media interactions using a coevolving latent space network with attractors model}
\author[Zhu {\it et al.}]{Xiaojing Zhu}
\address{Department of Mathematics and Statistics, Boston University, Boston, USA}
\author{Cantay Caliskan} 
\address{Goergen Institute for Data Science, University of Rochester, Rochester, NY}
\author{Dino P. Christenson}
\address{Department of Political Science, Washington University in St. Louis, St. Louis, USA}
\author{Konstantinos Spiliopoulos}
\address{Department of Mathematics and Statistics, Boston University, Boston, USA}
\author{Dylan Walker}
\address{Argyros School of Business and Economics, Chapman University, Orange, USA}
\author{Eric D. Kolaczyk}
\address{Department of Mathematics and Statistics, McGill University, Montreal, Canada}
\email{eric.kolaczyk@mcgill.ca}

\begin{document}
\maketitle

\begin{abstract}
  We develop a broadly applicable class of coevolving latent space network with attractors (CLSNA) models, where nodes represent individual social actors assumed to lie in an unknown latent space, edges represent the presence of a specified interaction between actors, and attractors are added in the latent level to capture the notion of attractive and repulsive forces. We apply the CLSNA models to understand the dynamics of partisan polarization on social media, where we expect Republicans and Democrats to increasingly interact with their own party and disengage with the opposing party. Using longitudinal social networks from the social media platforms Twitter and Reddit, we investigate the relative contributions of positive (attractive) and negative (repulsive) forces among political elites and the public, respectively. Our goals are to disentangle the positive and negative forces within and between parties and explore if and how they change over time. Our analysis confirms the existence of partisan polarization in social media interactions among both political elites and the public. Moreover, while positive partisanship is the driving force of interactions across the full periods of study for both the public and Democratic elites, negative partisanship has come to dominate Republican elites' interactions since the run-up to the 2016 presidential election.
\end{abstract}
\section{Introduction}

The increase in partisan polarization over the last five decades is one of the most consequential developments in American politics. Ample evidence from political science suggests that political elites, like members of Congress, increasingly disagree on policies 
\citep[e.g.,][]{McCartyetal2006,McCarty2019}.
Despite continued scholarly debate on the degree of ideological polarization among the public \citep{abramowitz2008polarization,fiorina2005culture}, there is little doubt today that substantial portions of the public are polarized in their affections towards the parties \citep{greenberg2005two,jacobson2007divider,bafumi2009new,iyengar2012affect,mason2015disrespectfully,huddy2015expressive}. 

Though the causal mechanisms of affective and ideological polarization remain active areas of research \citep[e.g.,][]{Levendusky2013,Mason2018,DiasLelkes2021}---as do their operationalizations \citep{Lelkes2016}---their documented manifestations are already many, including increases in party loyalty, straight-ticket voting and animosity towards candidates \citep{Jacobson2000,abramowitz2016rise,christenson2019bad}, as well as decreases in ambivalence, indecision and floating in elections \citep{smidt2017polarization}. Notably, considerations of the impact of polarization and political identities have spilled over onto less overtly political targets as well \citep{DellaPostaetal2015,HetheringtonWeiler2018}, such as marriage \citep{Alfordetal2011,Iyengaretal2012}, online dating and friendship \citep{HuberMalhotra2017,Bakshyetal2015}, where to live \citep{Bishop2009,ChristensonMakse2015,MummoloNall2017}, and who in the family to talk to \citep{ChenRohla2018}. In this work, we similarly devote our attention to exploring polarization in social life, but in the modern and evolving environment of social media.     

Social media is an increasingly popular and accessible forum for all kinds of discussion \citep{Brundidge2010,Holt2004}, yet the question of whether these discussions take place in silos of like-minded individuals is yet unresolved \citep{Barbera2020,Tuckeretal2018}. Scholarship has pointed to homophilic interactions online or \textit{echo chambers} \citep{HalberstamKnight2016,Bright2018,Bodrunovaetal2019, Yarchietal2021} as well as heterophilic ones \citep{Bruns2019, ZuiderveenBorgesiusetal2016},
with recent works providing nuanced insights through considerations of different platforms and over time change \citep{Yarchietal2021,WallerAnderson2021}. Indeed, social media platforms may facilitate polarization leading to greater fragmentation \citep{Bobok2016,HayatSamuel-Azran2017} via user selective exposure \citep{Settle2018} and algorithmic filtering towards user predispositions \citep{MutzMartin2001,Nahon2016}.  Scholarship finding heterogeneous engagements has noted an important role for online interactions in providing cross-cutting views, especially given polarization in the other spheres of modern political and social life \citep{Bakshyetal2015,Barberaetal2015,Barnidge2017,DuboisBlank2018,Eadyetal2019}. Such cross-cutting interactions may mitigate intergroup biases \citep{Bondetal2021,Marchal2021}, which, from the perspective of normative democratic theory, may be essential in reducing the discord in highly polarized political systems \citep{Mutz:2006,Pettigrew1998}.

With the magnitude and reach of partisan polarization at seemingly unprecedented levels in US politics and beyond, we seek to better understand its nature online. First and foremost, we ask: is there an increase in partisan polarized interactions on social media? And if so, is it due to increasing interactions with one’s own party or decreasing interactions with the other party---or both, as we might expect from social identity theory \citep{TajfelTurner1979}? Moreover, are these patterns consistent across Democrats and Republicans, and across elites and the public? Our goals then are both to check for partisan polarization's presence in social media interactions among the elites and the public, engaging the early debates on mass-elite polarization, and also to uncover the attractive and repulsive forces of polarization over time, which speaks to an emerging literature on negative partisanship. 

Originally noted in studies of multiparty voting \citep{rose1998negative,caruana2015power,medeiros2014forgotten}, the idea of negative partisanship is that negative evaluations of the out-group party (i.e., negative partisanship)---as opposed to positive evaluations of the in-group party (i.e., positive partisanship)---are dominant in political behavior and opinions.  Indeed, the movement in affective polarization has largely been in terms of increasingly negative feelings for the other party and its members, while feelings towards their own have remained fairly stable \citep{abramowitz2018negative,christenson2019bad}.
Such findings suggest an increased role for negative partisanship, which we evaluate by uncovering the relative strengths of attractive and repulsive forces of partisan in-group and out-group interactions on social media. 

We construct two longitudinal social networks from the social media platforms Twitter and Reddit to investigate partisan polarization and, more specifically, the contributions of positive (attractive) and negative (repulsive) partisanship among political elites and the public, respectively. In the following section we discuss in more detail each of the social media networks, but we note here that in both networks nodes represent individual social actors and edges represent the presence of a specified interaction among them. If the expectations from the aforementioned literature on partisan polarization above are correct then we should find social media users increasingly interacting with users from the same party, while disengaging from those of the other party. Of course, negative partisanship suggests a growing role for disengaging with the other party over interacting with their own. In addition, we recognize the possibility that these dynamics exhibit asymmetry, differing for elites and the public, as they might across the two parties. 

To test the expectations of polarization and negative partisanship in social media interactions, we develop a class of \textit{coevolving latent space network with attractors} (CLSNA) models in which both the links between nodes and certain characteristics (or attributes) of nodes evolve over time---each in a way that impacts the other. This class of models falls within the subclass of latent space models, of which static versions are now especially well-developed, and progress on dynamic versions has been made in recent years. Existing dynamic latent space network models \citep{sarkar2005dynamic, sewell2015analysis, sewell2015latent, sewell2016latent, sewell2017latent} dictate simply that latent characteristics evolve in time in a Markov fashion and then links between nodes exist with probabilities driven by node distance in the underlying latent space. 
In contrast, in our framework the temporal evolution of latent characteristics can depend on network connectivity---hence, a \textit{coevolving} network model.
Such a feedback mechanism is embedded into the model via the presence of attractors (a concept fundamental to dynamical systems) at the latent (i.e., unobserved) level. This class of coevolving network models is, to the best of our knowledge, novel, and the models are generall y applicable to any case of coevolutionary phenomena in social behaviors such as flocking and polarization.

For this application we define a two-group CLSNA model with two attractors mimicking attractive and repulsive forces, which is a specific version of the CLSNA model class. In this version of the model, each node in the network is assumed to fall into one of two groups with known labels, and their movements are influenced by their neighbors through specially defined attractors. We also account for persistence of links in the dynamic evolution of the network, which is a necessary control informed by social network theory. Finally, we define a change-point version of this model to allow for the potential of time-variation of parameters, where the network evolves according to one set of parameter values up to a given change-point, and another set of parameter values after the change-point.

This article is organized as follows: In the following section we provide an overview of the two network data sets used in this study through an exploratory analysis. We also introduce the statistical network model proposed in this work, including a discussion of model behaviors and parameter interpretation. Adopting a Bayesian perspective, we develop a Metropolis-Hastings (MH) within Gibbs MCMC framework for posterior inference. In the `Results' section we present the application of this model to the two data sets and quantitatively analyze the evolution of the key factors of social media interactions. Finally, we conclude the study with a discussion of the results and the model proposed.

\section{Network Data}
In this section, we introduce data used in this study. We construct and explore two different online longitudinal social interaction networks for evidence of polarization, one of the elite, via Twitter, and one of the public, via Reddit. Data and code are available at \texttt{https://github.com/KolaczykResearch/CLSNA-2Party-Polarization}.

\subsection{Twitter congressional hashtag networks} 
We collected tweets for every US congressperson with a handle from 2010 to 2020. This yielded 796 accounts, 843,907 tweets and 1,252,455 instances of hashtag sharing (after retweets and tweets with no hashtags are removed).
This data was used to construct a binary network for each year, wherein nodes correspond to sitting members of Congress and edges indicate that the number of common hashtags tweeted by both members of Congress that year was more than the average. We use the yearly average instead of a static threshold in constructing edges, because hashtag usage in Twitter by members of Congress increased substantially over the past decade, which allows our model to capture more meaningful interaction dynamics beyond increasing hashtag usage. The nodes for the resulting networks vary from year to year since some members of Congress were reelected, while others may have joined Twitter at a later stage, left early or both. In our analysis, we focus our analysis on the 207 members of Congress who served in office and stayed active on Twitter over the entire course of our study, among whom 131 are Democrats and 76 are Republicans. 

The choice of edge construction can be consequential in studies of polarization. While the approaches vary, here we follow the literature that has taken advantage of hashtag sharing \citep{borge-holthoefer_content_2015, magdy__2016, weber_secular_2013, darwish_quantifying_2019, bovet_validation_2018, kusen_politics_2018}. In doing so, we perceive of polarization in terms of \textit{indirect} engagement \citep[e.g.,][]{bruns_structural_2014, naaman_hip_2011, wilkinson_trending_2012, giglietto_hashtag_2017, bode_candidate_2015}---i.e., joining or avoiding the same topics of discussion. Relative to more direct engagement on social media, like retweeting or mentions, this approach provides a higher bar for polarization, since disengaging issues is an arguably more consequential activity---at least for members of Congress whose duty it is to address important topics---than just ceasing to engage with specific individuals. Our approach recognizes the key role of legislators in addressing important political topics regardless of party lines. In this framework, polarization is evinced when members discontinue engaging with the same topics and issues. 
We further discuss the edge construction in Section 1 of the SI Appendix.


\subsection{Reddit comment networks} 
In the case of Reddit, we collected the full data on submissions and comments since the site’s inception through pushshift.io. We focus on active Reddit users whose ideologies can be identified from their comments or flairs with declarative patterns, e.g., ``I am a Republican.''. We thereby selected 102 Republicans and 267 Democrats who made 1) at least one comment in each month during April 2015---March 2020, and 2) more than 50 comments in a year across political subreddits (e.g., `politics', `Libertarian', `PoliticalDiscussion', `Conservative', etc.). We then constructed longitudinal binary networks on these 369 active users for each one-year period based on their interactions in comments, wherein an edge between two active users indicates that they commented on the same submission. Like the Twitter analysis, the edge here is based on engaging the same submission or topic---and not necessarily direct engagement among two or more users. Because average commenting rates in Reddit did not change substantially over time, we use a static threshold in defining edges.

\subsection{Exploratory Analysis}
\begin{figure}[tbhp]
\centering 
\includegraphics[width=0.5\linewidth]{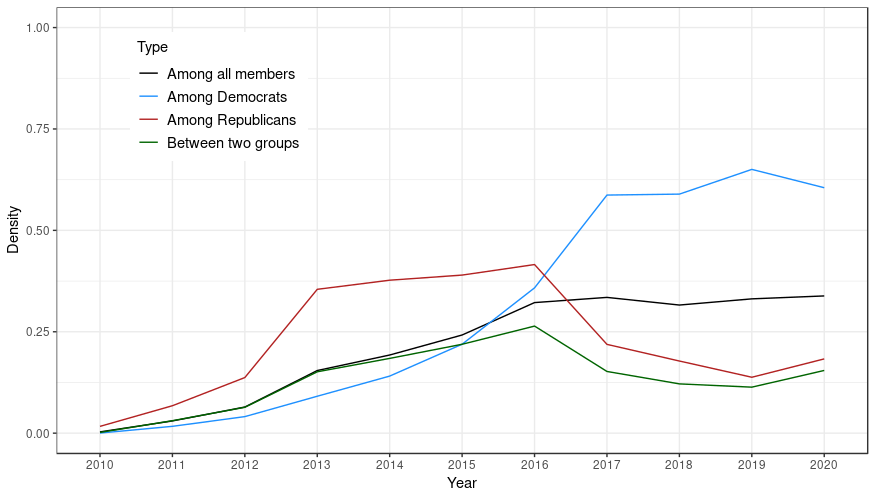}
\caption{Twitter congressional hashtag networks. Four types of edge densities over years (inter-party, intra-party and overall) for constructed networks on 207 members of Congress who served in office and stayed active in Twitter over the entire course of our study from 2010 to 2020.}
\label{twitter_den}
\end{figure}

\begin{figure}[tbhp]
\centering 
\includegraphics[width=0.5\linewidth]{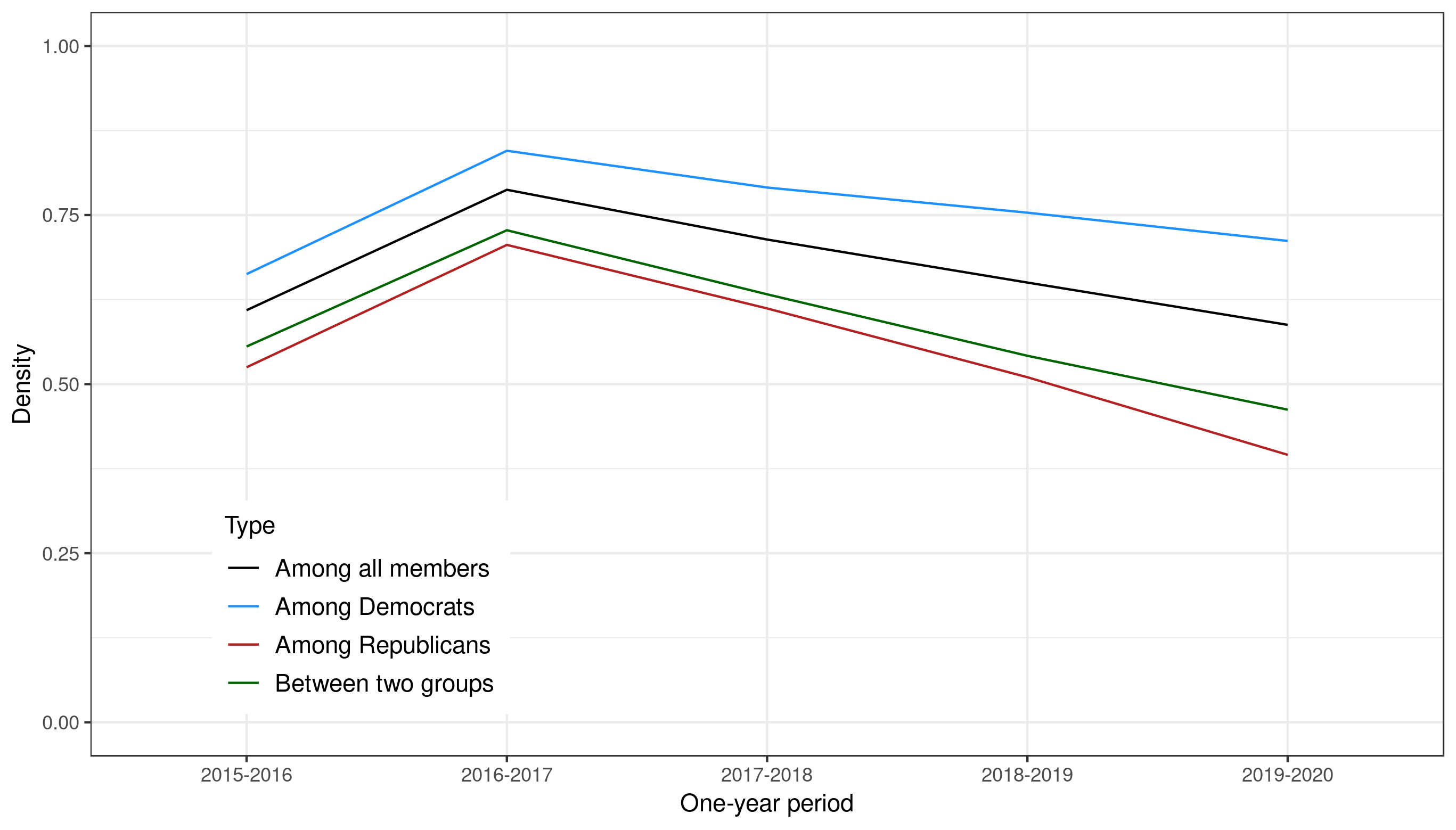}
\caption{Reddit comment networks. Four types of densities over years (inter-party, intra-party and overall) for constructed network on 369 active Reddit users identified as either Democrats or Republicans.}
\label{reddit_den}
\end{figure}

Figure \ref{twitter_den} and Figure \ref{reddit_den} show plots of the density of edges within and between Democrats and Republicans as they evolve over time for the Twitter and Reddit networks, respectively. In Figure \ref{twitter_den} for Twitter congressional hashtag networks, while initially growing together in density, over the last four years, we notice divergent trends among the subsets of Democrats and Republicans, with Democrats increasing in their social media connections to each other and Republicans decreasing. While connections have increased overall, the presidential election year of 2016 brought about a drop in inter-party connections. In Figure \ref{reddit_den} for Reddit comment networks, while intra-party connections for the two parties share similar trends, with both of them initially increasing then decreasing, they differ in the extent of drop over the last three years, with more interaction ties dissolved among the subset of Republicans than Democrats. The inter-party connections also have decreased following the election year of 2016. 

This is the first hint that the evolving polarization of the elite and general public begs a more complex interpretation than is usually brought to bear. Of course, this is only descriptive, and thus we turn to inferential methods with our proposed two-group CLSNA model with attraction and repulsion.

\section{A Two-group Coevolving Latent Space Network with Attractors (CLSNA) Model with Attraction and Repulsion for Polarization}
In this section, we outline our specific statistical modeling and inference approach. 
\subsection{Model Definition}
Let $G_t=(V_t,E_t)$ be a network evolving in (discrete) time $t$, with vertex set $V_t$ and edge set $E_t$.  For simplicity, assume that $V_t\equiv V$ is fixed over time, of order $N = |V|$.  Let $Y_t$ be the (random) adjacency matrix at time $t$ corresponding to $G_t$. Throughout the paper, we use capitals to denote random variables and lower-cases to denote the realizations of them. We assume data come in the form of time series of adjacency matrices $\{y_t: t=1,\cdots, T\}$, where $y_{t, ij}=1$ if there is an edge between node $i$ and node $j$ at time $t$ and $0$ otherwise. 

To model the dynamic evolution of networks in connection with our expectations of polarized online interactions, we use the latent space approach and add attractors in the latent level to capture the notions of attractive and repulsive forces at the heart of negative partisanship. Let $\bm z_{t,i} \in \mathbb{R}^p$ be a time-indexed latent (i.e., unobserved) position for node $i$ in $p$-dimensional Euclidean space, and $\bm z_t = \{z_{t,i}\}$.  Assume that each of the $N$ nodes of the network falls into one of two groups, i.e.,  Democratic and Republican, with known node label $\pi(i) \in \mathcal{C}$ for node $i$, where $\mathcal{C} = \{1, 2\}$ is the set of group labels. Formally, we define our model as follows:
\begin{eqnarray}
Y_{t,ij}\, |\, p_{t,ij} & \sim & \hbox{Bernoulli}(p_{t,ij}) 
\label{eq:bernoulli} \\
\hbox{logit}(p_{t,ij}) & = & \alpha + \delta Y_{t-1,ij} - s(\bm z_{t,i}, \bm z_{t,j}) \label{eq:logit} \\
\bm Z_{t,i}\, |\, \bm Z_{t-1,i}  = \bm z_{t-1,i} 
& \sim &
\hbox{Normal} \, \big ( \, \bm z_{t-1,i} + 
\nonumber\\
& & 
{ \gamma^{w}_{\pi(i)}} A^w_i\left(\bm z_{t-1},Y_{t-1} \right) +  
\label{eq:z1}\\
& & 
\gamma^b A^b_i\left(\bm z_{t-1},Y_{t-1} \right) 
\, ,\, \sigma^2 I_p \, \big) 
\nonumber \enskip ,
\end{eqnarray}
with initial distribution at time $t=1$,
\begin{eqnarray}
\bm Z_{1,i}  & \sim & \hbox{Normal}\left( \bm 0, \, \tau^2 I_p\right)  
\label{eq:z_init}\\
\hbox{logit}(p_{1,ij}) & = & \alpha - s(\bm z_{1,i}, \bm z_{1,j}) 
 \label{eq:logit_init} \enskip .
\end{eqnarray}
Here $s(\cdot, \cdot)$ is a similarity function, and $A_i^w$ and $A_i^b$ are the two attractor functions for node $i$ in $Y_{t-1}$. Specifically, we set $s(z_{t,i}, z_{t, j}) = ||z_{t,i}-z_{t, j}||_2$, and define the two attractors for node $i$ as follows, 
\begin{equation}
A_i^w(\bm z_{t-1},Y_{t-1}) = \bar{\bm z}^1_{t-1,i}- \bm z_{t-1,i}, \,
\bar{\bm z}^1_{t-1,i} = \frac{1}{|S_1^{(i)}|} \sum_{j\in S_1^{(i)}} \bm z_{t-1,j}
\label{eq:A1}
\end{equation}
\begin{equation}
A_i^b(\bm z_{t-1},Y_{t-1}) = \bar{\bm z}^2_{t-1,i}- \bm z_{t-1,i}, \,
\bar{\bm z}^2_{t-1,i} = \frac{1}{|S_2^{(i)}|} \sum_{j\in S_2^{(i)}} \bm z_{t-1,j}
\label{eq:A2}
\end{equation}
which are the discrepancies of $\bm z_{t-1,i}$ from two local averages at time $t-1$. These latter are the average of latent values of nodes in the following two sets, informed by a combination of group membership, and network connectivity:
\begin{enumerate}
\item 
$ S_1^{(i)} = \{ j \in \mathcal{N} \setminus i \,|\, Y_{ij} = 1, \pi(i) = \pi(j) \}$, neighbors of node $i$ in the same group
 \item 
 $  S_2^{(i)} = \{ j \in \mathcal{N} \setminus i \,|\, Y_{ij} = 1, \pi(i) \neq \pi(j) \}$, neighbors of node $i$ in a different group.
\end{enumerate}
When $S_1^{(i)}$ or $S_2^{(i)}$ or both are empty, we set $A_i^w(\bm z_{t-1},Y_{t-1})=0$, or $A_i^b(\bm z_{t-1},Y_{t-1})=0$ or both to be zero. This means that when node $i$ has no neighbors in a certain group, or no neighbors at all, there is no attraction from that group, or no attraction at all.

In this proposed model, we assume that each node lies in a $p$-dimensional Euclidean latent space, and the smaller the distance between two nodes in the latent space, the greater their probability of being connected, as in \eqref{eq:bernoulli}, \eqref{eq:logit}. The expressions in Eqs. (\ref{eq:A1}) and (\ref{eq:A2}) capture the discrepancy between the current latent position of node $i$ and the average of that of its current neighbors in groups $1$ and $2$, respectively. The corresponding parameters $\gamma^w_1, \gamma^w_2,$ and $\gamma^b$ represent attractive/repulsive forces, as we discuss below.

\begin{figure}[tbhp]
\begin{center}
\vspace{-0.1in}
\hspace{-2in}
\begin{tikzpicture}[node distance = 2cm, thick]
        \node (1) {$Y_{t-1,ij}$};
        \node (2) [right of=1] {$Y_{t,ij}$};
        \node (3) [below of=1] {$Z_{t-1}$};
        \node (4) [below of=2] {$Z_{t}$};
       \draw[->] (3) -- (1);
        \draw[->] (4) -- (2);
        \draw[->] (3) -- (4);
\hspace{2.0in}
        \node (1) {$Y_{t-1,ij}$};
        \node (2) [right of=1] {$Y_{t,ij}$};
        \node (3) [below of=1] {$Z_{t-1}$};
        \node (4) [below of=2] {$Z_{t}$};
       \draw[->] (3) -- (1);
        \draw[->] (4) -- (2);
        \draw[->] (3) -- (4);
        \begin{scope}[color=red]
        \draw[->] (1) -- (2);
        \end{scope}
        \begin{scope}[color=blue]
        \draw[->] (1) -- (4);
        \end{scope}
\end{tikzpicture} 
\vspace{-0.1in}
\end{center}
\caption{Graphical model representations of existing dynamic latent space network models (left), and proposed CLSNA (right) with both node attraction (blue arrow) and edge persistence (red arrow).}
\label{graphical_rep}
\end{figure}
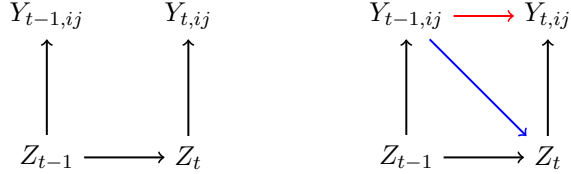

In contrast to the existing dynamic latent space network models \citep{sarkar2005dynamic, sewell2015analysis, sewell2015latent, sewell2016latent, sewell2017latent} where the latent process is assumed to evolve over time in a Markov fashion with transition distribution, e.g., $\bm Z_{t,i}|\bm Z_{t-1,i}  = \bm z_{t-1,i} \sim
\hbox{Normal}( \bm z_{t-1,i}, \sigma^2 I_p \, ) $, and thus to drive evolution of the networks, as illustrated in Figure \ref{graphical_rep} (left), our CLSNA model allows the network connectivity to enter the temporal evolution of latent positions in the form of attractors, as illustrated by the blue arrow in Figure \ref{graphical_rep} (right). Specifically, in our model the evolution of latent positions for each node $i$ from $t-1$ to $t$ is modeled by the normal transition distribution in Eq. (\ref{eq:z1}), the mean vector of which depends not only on the latent position of itself at time $t-1$, but also on the two local averages, one from its neighbors in the same group, the other from its neighbors in a different group, as captured in \eqref{eq:A1} and \eqref{eq:A2}. This is an important aspect of our model since it quantifies propensity for attraction/repulsion within/between two groups, and can help us understand how polarization/flocking and interaction co-evolve. Strength of attraction/repulsion toward local averages is therefore summarized by the attractor functions and the associated parameters, the details of which are discussed in the later sections. 

We also include an effect for edge persistence, as illustrated by the red arrow in Figure \ref{graphical_rep} (right), which is a necessary control informed by social network theory. $\delta$ captures the impact of having an edge at time $t-1$ on whether or not there is an edge at time $t$.  For $\delta>0$, the probability of an edge at time $t$ will be increased when one exists already at time $t-1$, and hence the model explicitly captures a notion of edge persistence.
 
\subsection{Model Behavior and Parameter Interpretation} 
Our model incorporates a level of baseline connectivity ($\alpha$), edge persistence ($\delta$), two separate within-group node attraction for the two groups ($\gamma_1^w$ and $\gamma_2^w$, respectively), between-group node attraction ($\gamma^b$), and a measure of volatility ($\tau^2$ initially, and $\sigma^2$ for $t>1$). A rich set of behaviors can be generated by varying these parameters. The three attraction parameters $\gamma_1^w$, $\gamma_2^w$ and $\gamma^b$ are of particular interest, in that by varying the sign they allow for the possibility of different combinations of attraction and/or repulsion in the evolution of the latent positions. The sign of these parameters encodes the direction of these forces -- a positive sign indicates latent positions being pulled toward the direction of local averages, aka attraction, while a negative sign indicates being pushed toward the opposite direction, aka repulsion. For example, when $\gamma_{1}^w, \gamma_{2}^w, \gamma^b >0$, we can interpret this as two groups flocking together, while for $\gamma_{1}^w, \gamma_{2}^w > 0$ but $\gamma^b <0$, the two groups are flocking separately--- that is, we have a notion of polarization. 

\begin{figure}[tbh]
    \centering
    \includegraphics[width=0.35\textwidth]{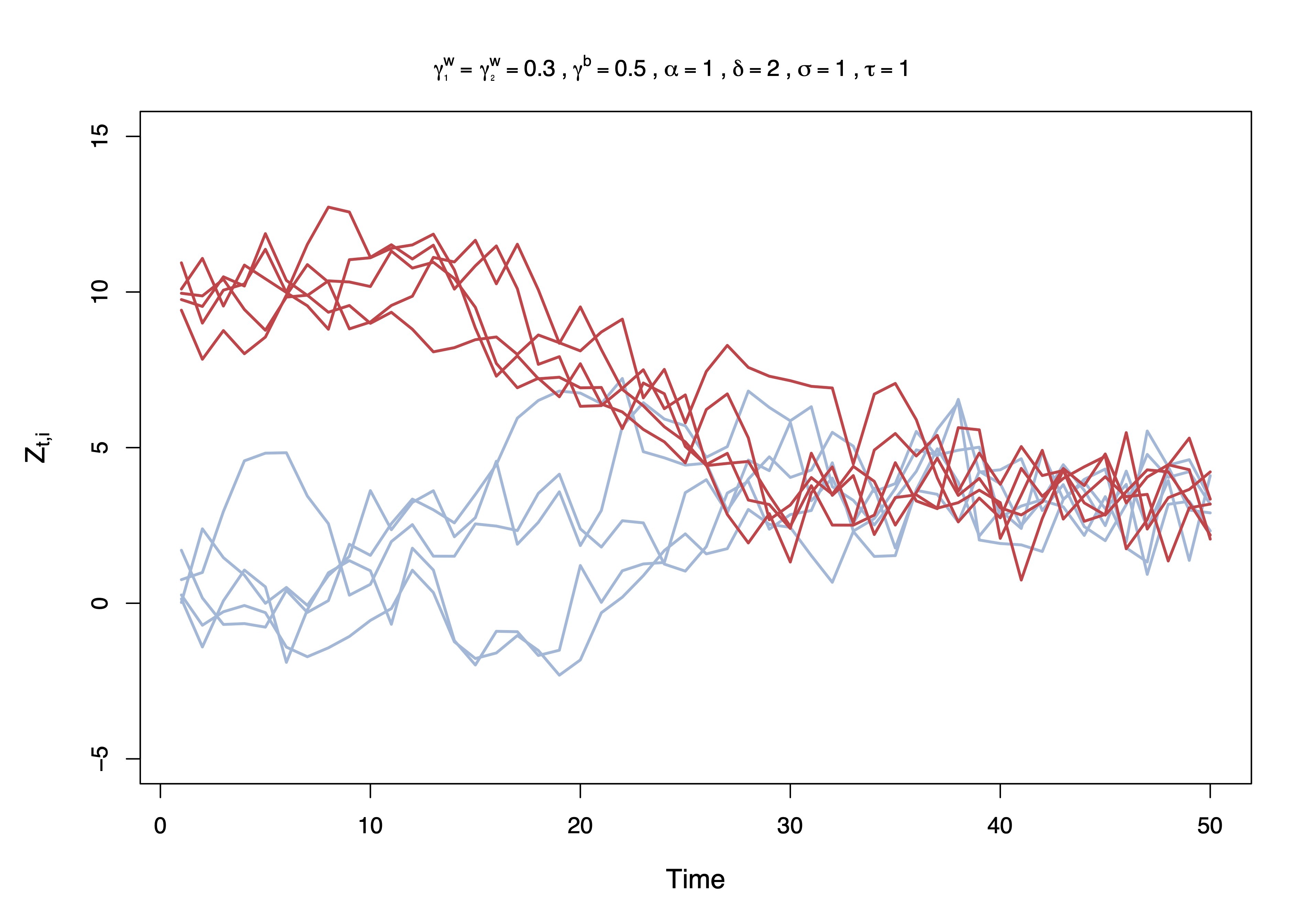}
    \includegraphics[width=0.35\textwidth]{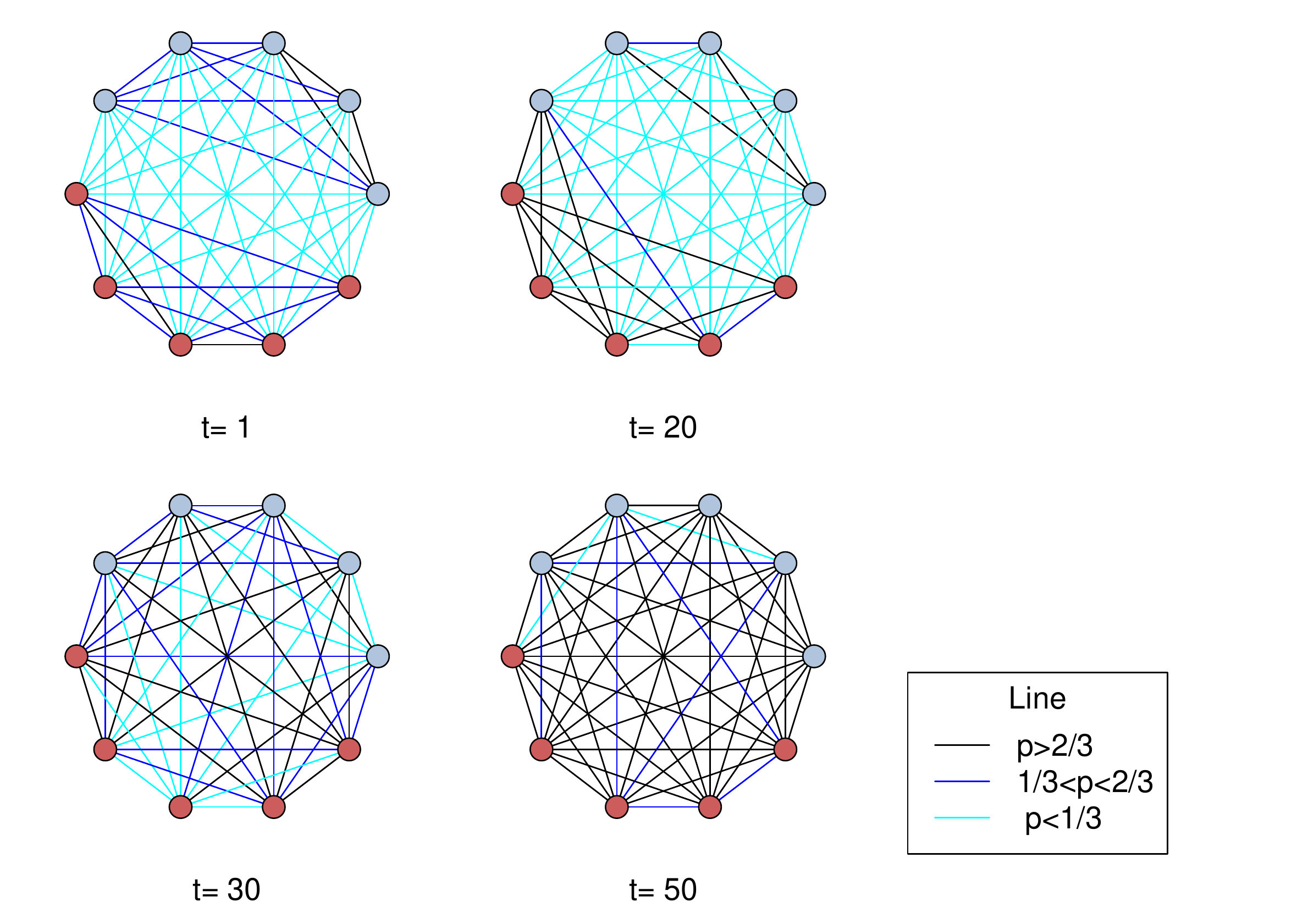}
    \includegraphics[width=0.35\textwidth]{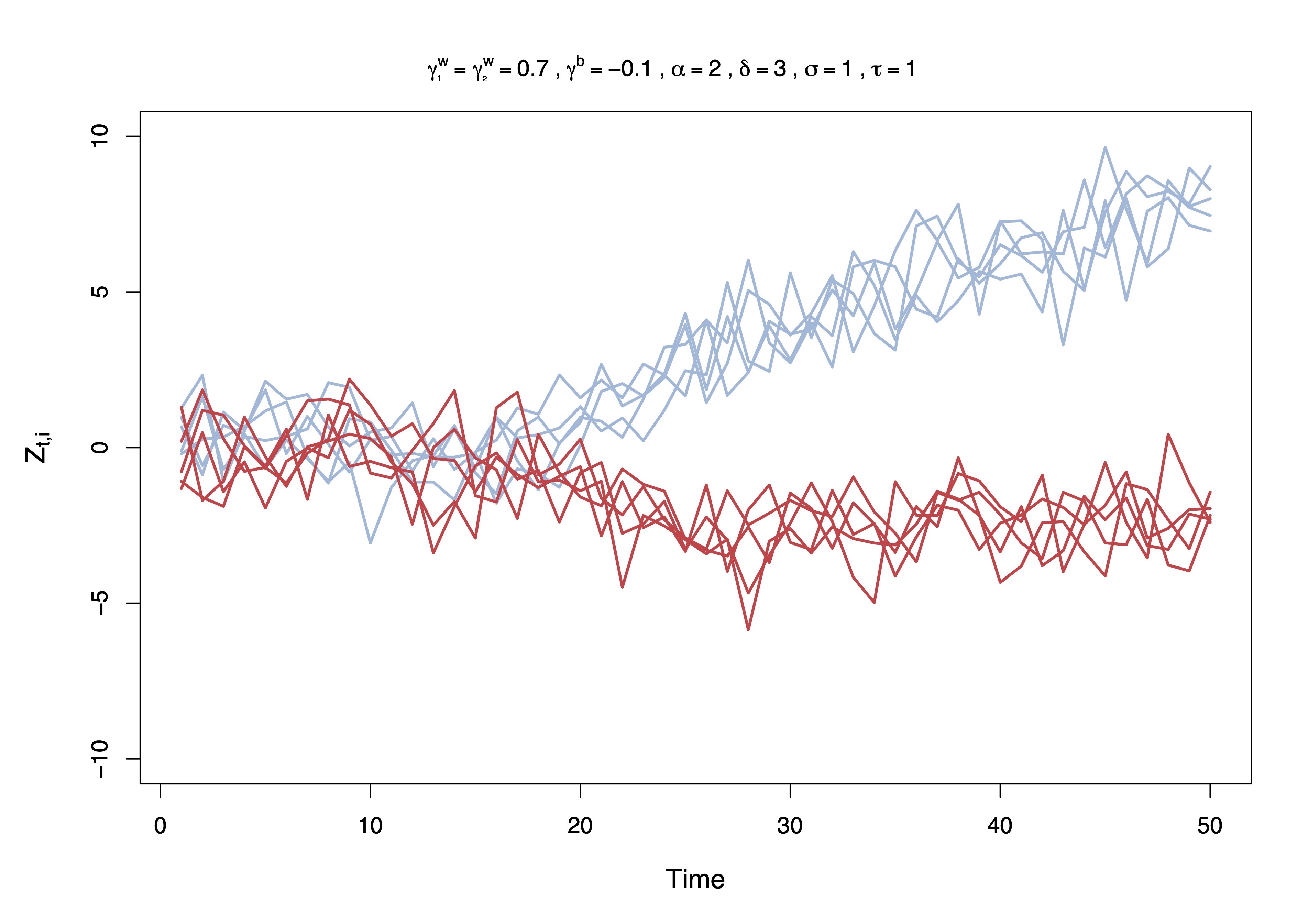}
    \includegraphics[width=0.35\textwidth]{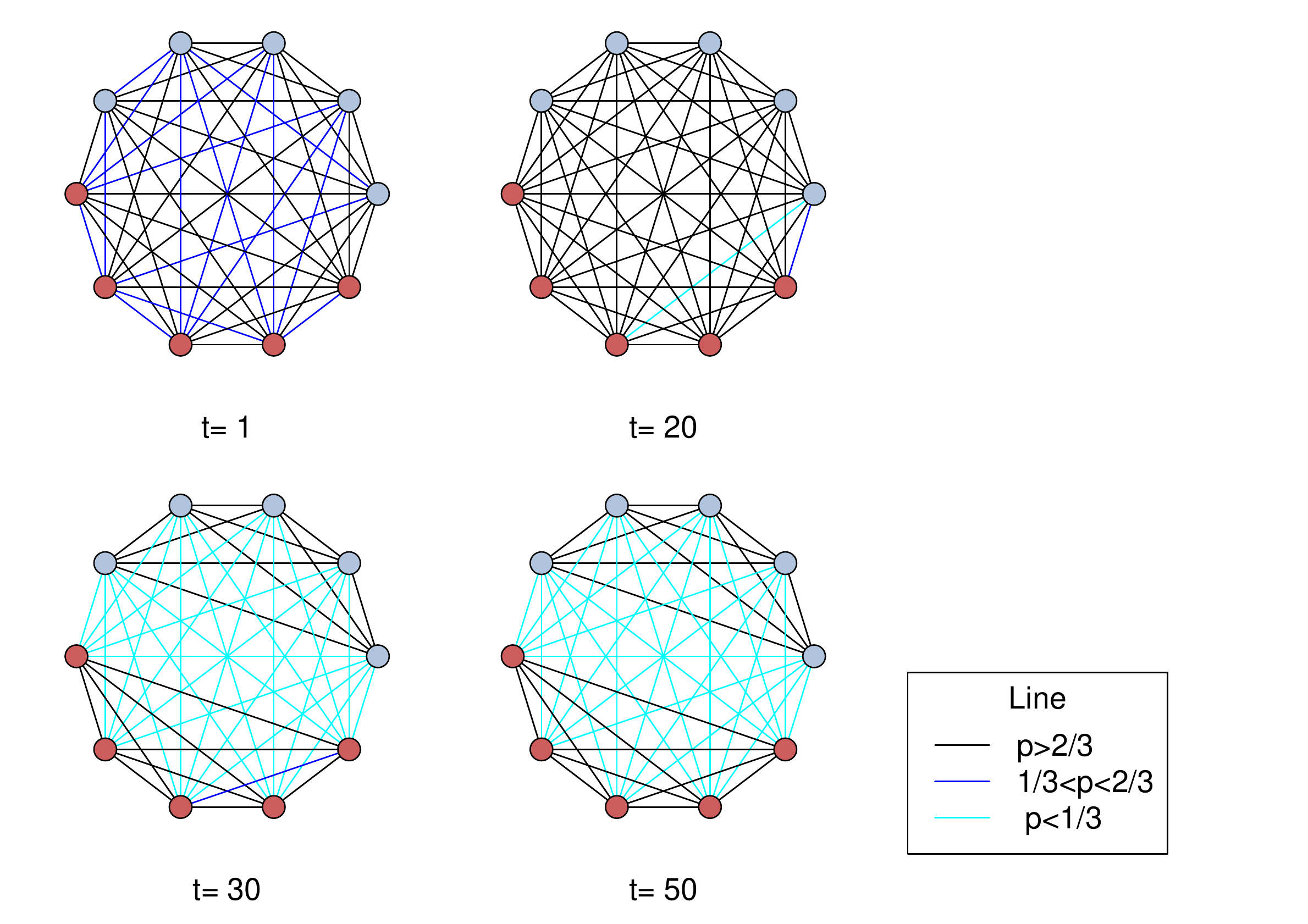}
    \caption{Top: Flocking setting ($\gamma_1^w, \gamma_2^w, \gamma^b > 0$). Bottom: Polarization setting ($\gamma_1^w, \gamma_2^w>0, \gamma^b <0$). Left: Trajectories of latent positions by time for two groups in 1-dimensional space. Right: Snapshots of the evolution of the probability of connections between different nodes in the graph.}
\label{fig-behavior}
\end{figure}

In Figure \ref{fig-behavior}, we illustrate the behavior of latent positions and network connectivity in simulated models for the two scenarios, one reflecting two group flocking, and the other, polarization among the same two groups. For convenience of visualization, the latent space is taken to be one-dimensional. We can see that initialized with different latent positions, the time courses for positions of the $N=10$ nodes in this network cluster together under flocking. But initialized together, they diverge into two clusters under polarization. At the same time, while the network becomes ever more densely connected over time under flocking, it evolves towards two fully connected subgraphs under polarization. 

One interesting observation from our simulation studies is that setting $\gamma_1^w, \gamma_2^w$ too large could lead to divergent model behaviors: we lose the attraction effect since nodes are being pulled too far away from the local averages, although the repulsion effect seems to be well-behaved for large magnitude of $\gamma^b$. This indicates that there is an interesting balancing relation between the different parameters of the model that warrants further theoretical investigation. Such an undertaking is beyond the scope of this work, but it certainly is an interesting avenue for further research. 


\subsection{Quantifying the Extent of Inter-/Intra-party Attraction/Repulsion and Edge Persistence} 

While the sign of each attraction parameters $\gamma_{1}^w, \gamma_{2}^w, \gamma^b$ encodes attraction (positive sign) or repulsion (negative sign) within group 1, within group 2 and between the two groups, respectively, the absolute value of these parameters can be used to quantify the extent of inter-/intra-party attraction/repulsion. Similarly, the value of $\delta$ can be used to quantify the relative importance of edge persistence. Accordingly, we take the magnitude of inter-party repulsion as a measurement for negative partisanship, and that of intra-party attractions for positive partisanship.

We can answer an array of questions regarding the nature of polarization online and, more specifically, negative partisanship by investigating the values of these parameters. For example, does the phenomenon of partisan polarization occur in our Twitter and Reddit data? If so, is repulsion to the other party or engagement with one's own the dominant factor driving the interactions among elites and the public online? These questions can be answered respectively by assessing whether $\gamma_b<0$; and by comparing the values of $|\gamma_1^w|$, $|\gamma_2^w|$ with $|\gamma_b|$. Essentially, positive (negative) $\gamma_b$ indicates positive (negative) interaction towards the out-party, and $|\gamma_b|$ measures the extent of out-party disengagement. Similarly for $\gamma_1^w$, $\gamma_2^w$, the signs encode in-party positive/negative interaction, and the magnitude encode the extent of in-party engagement or disengagement.

\subsection{Bayesian Inference}
The parameters in our model are natural and interpretable candidates for statistical inference. Given the hierarchical nature of our model, Bayesian inference based on appropriate interrogation of the posterior distribution makes sense. That is, given an observed network time series $\{G_t\}_{t=1}^T$ or, more specifically, a time series of the corresponding adjacency matrices $\{Y_t\}_{t=1}^T$, we can make inference of the latent positions $\bm Z_t$ and model parameters $\bm \theta = (\alpha, \delta, \gamma^w_1, \gamma^w_2, \gamma^b, \tau^2, \sigma^2)$ based on the posteriors $p(\bm Z_{1:T}, \bm \theta \,|\, Y_{1:T})$. A closed-form expression for this distribution is not available, but we can use Markov chain Monte Carlo. We have implemented an adaptive Metropolis-Hastings (MH) within Gibbs MCMC scheme for posterior sampling. The implementation is non-trivial, as certain issues of scaling (regarding the volatility parameters $\sigma^2$ and latent positions) and rotational invariance (in the latent space) must be resolved. 
To be specific, we fix $\sigma^2=1$ to solve the issue of lack of identifiability of the model. Besides serving as a solution to a technical problem in MCMC where two conditional distributions are not identifiable, it essentially enforces the model, specifically $\hbox{logit}(p_{t,ij})$, to be scale-invariant with respect to $\sigma^2$. By fixing $\sigma^2=1$, we set a baseline reference for $\hbox{logit}(p_{t,ij})$. Details of the MCMC algorithm and the rationale of fixing $\sigma^2=1$ are provided in Section 2 and 3 of the SI Appendix.

\section{Results}
In this section, we fit our model to both Twitter data and Reddit data, with a latent space of dimension $p=2$, and present the estimates for model parameters and latent positions. The choice of two dimensions is consistent with DW-NOMINATE, one of the most popular established ideal point models of congressional ideology, for which 2 dimensions explain up to 90\% of variation in roll call voting \citep{poole2001}. To evaluate how well the model explains the data used to fit the model, we obtain the in-sample edge predictions by plugging the estimates into the linkage probability function and compute the AUC (area under the ROC curve) \citep{hanley1982meaning}. 

\subsection{Twitter Data Analysis}
We first fit our model with time-invariant parameters to the whole sequence of longitudinal networks in Twitter from year 2010 to 2020. The AUC values computed at each year are all above $0.976$, and the overall AUC value computed across all times is $0.988$, providing evidence that our model fits the data very well. 

The summary statistics for the posterior distribution of model parameters are provided in Table \ref{tab_para_twitter}. The edge persistence coefficient indicates that the log-odds that an edge appears increase by $1.5$ if the same edge appeared in the previous time frame. The between-group coefficient is $-0.155$, demonstrating polarization across the sets of Republican and Democratic members of Congress. Additionally, the within group coefficient is $0.493$ for Democrats, and $0.105$ for Republicans, which means that while they have moved away from one another, they generally flocked to their own. Moreover, the Democratic members have a higher extent of intra-party attraction on Twitter, meaning that
for Democratic members the interactions with their own party were stronger than those for Republican members.
Comparing the magnitude of between-group coefficients to the two within-group coefficients, we can see that for Democrats the interactions toward their own party were stronger than the disengagement with the other party, while for Republicans negative partisanship was dominant.

\begin{figure}[tbh]
\centering
\includegraphics[width=0.55\linewidth]{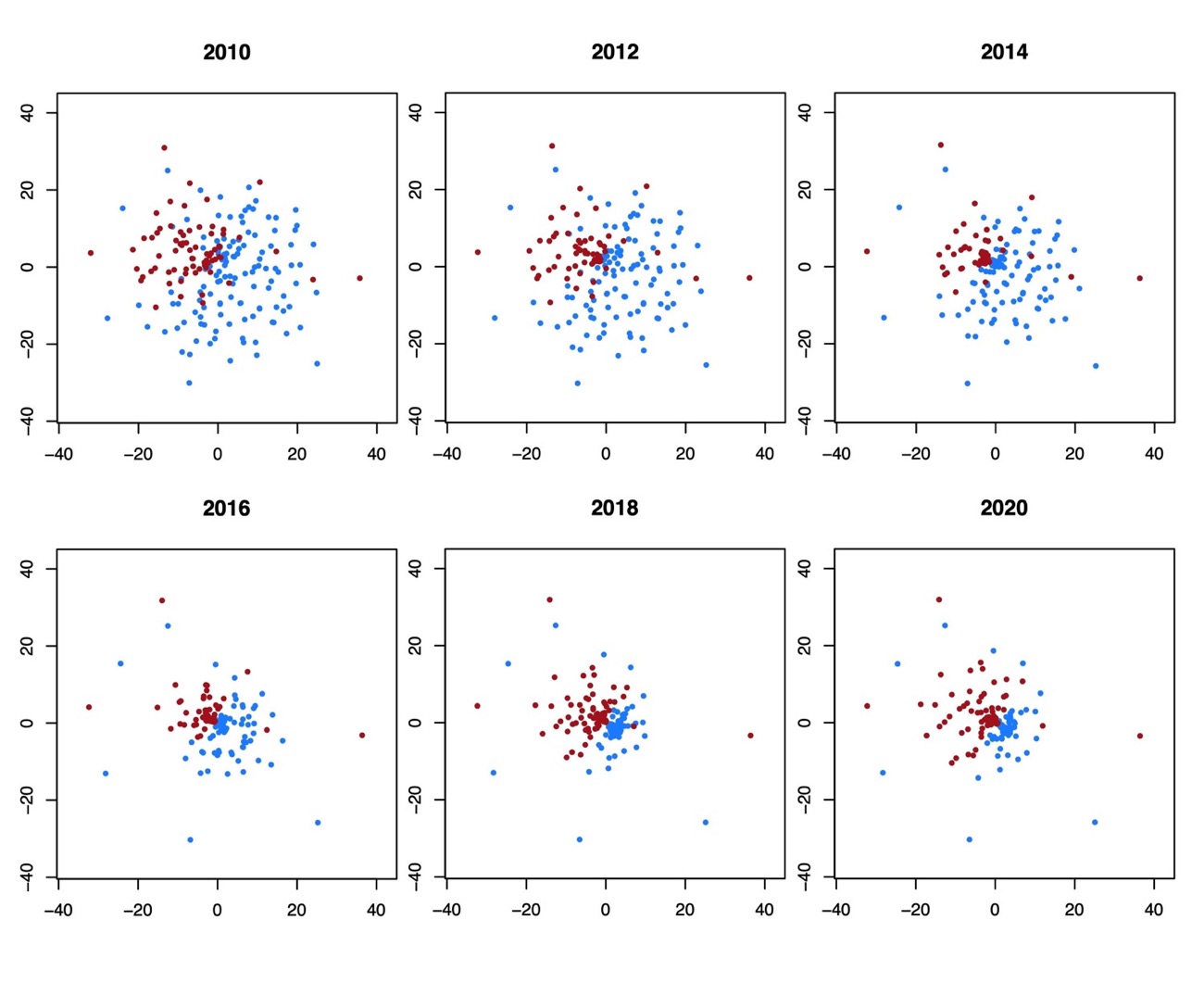}
\includegraphics[width=0.55\linewidth]{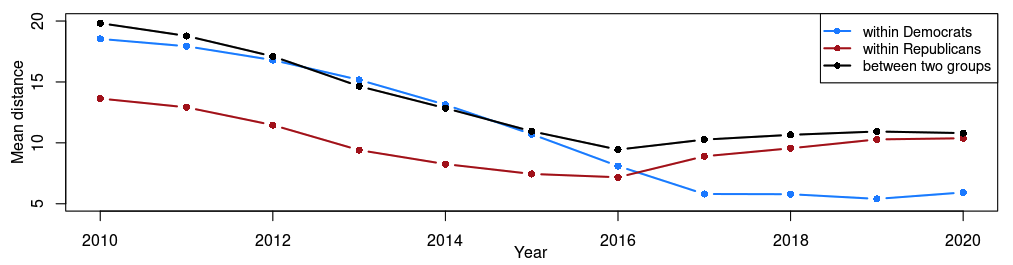}
\caption{\textit{Top}: Posterior means of latent positions for Twitter congressional hashtag networks (displaying only a subset of years). Blue-Democrats, Red-Republicans. Blue dots get closer over time, while red dots get closer first and then start spreading after 2016. \textit{Bottom}: Mean latent distances within each group and between groups.}
\label{fig_latent_twitter}
\end{figure}

Figure \ref{fig_latent_twitter} shows the posterior means of latent positions for each member of Congress in the Twitter hashtag networks. The dynamics of the clustering of latent positions exhibits a clear consistency with the evolution of within/between party edge densities seen in Figure \ref{twitter_den}, with Democratic members of Congress (blue dots) tending to converge over time, while Republican members of Congress (red dots) initially converge then disperse following the presidential election year of 2016. The inflection around 2016 seen in the bottom panel of Figure \ref{fig_latent_twitter} suggests that dynamics driving partisanship have changed. 

\begin{table}
\caption{Summary statistics for the posterior distribution of parameters using the whole sequence of Twitter networks from 2010 to 2020.}
\begin{threeparttable}
\centering
\begin{tabular}{*{7}{c}}
 &$\hat \tau^2$& $\hat \alpha$ & $\hat \delta$ & $\hat \gamma_1^w$ & $\hat \gamma_2^w$ & $\hat \gamma^b$\\\hline
Mean &125.763 & 2.809 & 1.500 & 0.493 & 0.105 & -0.155 \\
SD & 9.268 & 0.022 & 0.018 & 0.026 & 0.025 & 0.014 \\ 
2.5\% Quantile & 108.881& 2.766 & 1.464 & 0.444 & 0.055 & -0.183 \\ 
97.5\% Quantile & 145.193& 2.850 & 1.537 & 0.543 & 0.153 & -0.127 \\ 
\hline
\end{tabular}
\begin{tablenotes}
\item 1-Democrats, 2-Republicans. These results indicate presence of edge persistence ($\hat \delta > 0$), higher within-group attraction for Democrats than Republicans ($\hat \gamma_1^w > \hat \gamma_2^w$ (mean difference = .388, SE = .035)), presence of between-group repulsion ($\hat \gamma^b <0 $), and some evidence for greater magnitude of between-group repulsion than within-group attraction for Republicans ($|\hat \gamma^b| > |\hat \gamma_2^w|$ (mean difference = .050, SE = .022)), the opposite for Democrats ($|\hat \gamma^b| < |\hat \gamma_1^w|$ (mean difference = -.338, SE = .027)). 
\end{tablenotes}
\end{threeparttable}
\label{tab_para_twitter}
\end{table}

So far, we have seen that the CLSNA model is quite powerful in terms of revealing polarization in social network interactions, as well as disentangling and quantifying the two sides of polarization: positive and negative partisanship. These results motivate questions about the dynamics of the relationships uncovered above. In particular, given a host of major political, social and economic events over the past decade, can our model help us pinpoint changes in polarization and edge persistence over this period?

In order to confirm and quantify change in attraction, repulsion and edge persistence, we fit a series of models that allow a single change-point to vary from 2012 to 2019. Specifically, for each choice of change-point, we parameterized our model separately within the two corresponding subperiods of time, thus obtaining a set of parameter values up to the given change-point, and similarly another set of parameter values after the change-point. The resulting eight fitted models with different change-points were compared through deviance information criteria (DIC) \citep{spiegelhalter2002bayesian}, and the one with the lowest DIC value was selected (the DIC values for all competing models are provided in Section 7 of the SI Appendix). Our modeling identified 2015 as the year in which the network relationships changed the most.  The AUC values computed at each year for this model are all above $0.976$, and the overall AUC value computed across all times is $0.988$. From this model, we obtain the posterior means and 95\% credible intervals for the parameters $\delta$, $\gamma_1^w$, $\gamma_2^w$ and $\gamma^b$, for each of the two time periods 2010-2014 and 2015-2020.

\begin{figure}[tbh]
\centering
\includegraphics[width=0.55\linewidth]{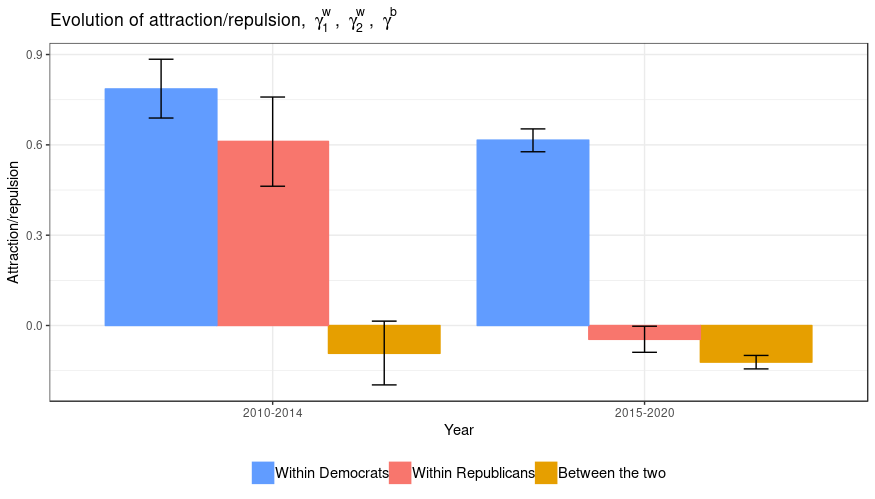}
\caption{Evolution of posterior means and 95\% CI for within-group attraction/repulsion $\gamma_1^w$ for Democrats, $\gamma_2^w$ for Republicans, and between-group attraction/repulsion $\gamma^b$ in Twitter congressional hashtag networks, the values of which at each time period in the horizontal axis are obtained by fitting the model using separately parameterized networks in the corresponding time periods. Polarization between two parties appeared in both time periods. The within-group coefficients for both  parties decreased. In the second time period, the between-group repulsion is greater than the repulsion within Republicans ($|\hat \gamma^b| > |\hat \gamma_2^w|$ (mean difference = .075, SE = .028)); i.e., for Republican members of Congress negative partisanship began to dominate in the second period, as disengagement with Democrats started to overcome engagements with their own. For Democratic members of Congress, positive partisanship dominates the entire period ($|\hat \gamma^b| < |\hat \gamma_1^w|$).}
\label{fig_twitter_gamma}
\end{figure}

Edge persistence appears to be fairly stable in the two time periods (shown in Figure 1 in the SI Appendix). Figure \ref{fig_twitter_gamma} illustrates the evolution of within-group attraction/repulsion $\gamma_1^w$ for Democrats, $\gamma_2^w$ for Republicans, and between-group attraction/repulsion $\gamma^b$. The between-group coefficient $\gamma^b$ (yellow bars) is negative in both time periods, although its magnitude increases a bit (mean increase = .027, SE= .051, P(increase$>$0)=0.701) in the second time period from 2015 to 2020. This suggests polarization across the sets of Republican and Democratic members of Congress appeared throughout the past decade, with some indication that it started to rise in 2015.

The within-group attraction coefficients for Democrats (blue bars) remain fairly large for the two time periods, albeit with a slight drop (mean decrease = .170, SE = .053) in the second period, while those for Republicans (red bars) exhibit a steeper downward trend falling from positive to negative. That is, for Democratic members of Congress the interactions with their own party have remained fairly strong, even after 2015 and during unified government under the Trump administration. For Republicans, however, the interactions with their own party are weaker (mean difference = .174, SE = .087) than they are for Democrats prior to 2015. Most intriguing, perhaps, Republican in-party interactions became negative in 2015.  That is, during the Trump administration, Republican members of Congress not only remained disengaged with Democrats, but also began to disengage with their own---i.e., a decrease in strength of in-group interaction.   
 
By comparing the magnitude of within-group coefficients (blue or red bars) with between-group coefficients (yellow bars), we find that for Democratic members of Congress positive partisanship dominated the entire time period of study from 2010 to 2020. Democrats interactions with their own was a binding feature of their polarization. In contrast, for Republican members of Congress disengagement with Democrats started to dominate interactions with their own in 2015 (mean difference = .075, SE = .028). Indeed, Republicans' interactions with their own party fell as well. Although not explicitly tested, the evidence here suggests that Trump's extensive appearance in social media and candidacy declaration in 2015 either caused or occurred in conjunction with the take-over of negative partisanship for Republican members of Congress. Whether due to long evolving attitudes within members of Congress or concurrent political trends, or simply reactions to the new presidential candidate, from 2015 to 2020 Republican members of Congress were defining their online partisanship more in terms of their disengagement with Democrats than engagement with other Republicans.

We have so far restricted the analysis to only one change-point. We could of course continue this analysis with more than one change-point. For example, we have run a series of models with two change-points chosen between 2012 and 2019, and selected that model with the lowest DIC value, which places change-points at year 2014 and 2019. However, the relative improvement of this model over that with a single change-point is quite modest. See Section 7 of the SI Appendix for details. 

\subsection{Reddit Data Analysis}
In this section we carry out the same line of analysis on the Reddit comment networks for the public. Recall that the Reddit data collected are from April 2015 to March 2020. Each network constructed represents the interaction during a one-year period from April of a given year to March the year after, and hence there are in total five networks constructed. We fit a model with a single set of parameters for the entire 2015-2020 period.  In addition, we fit a series of models with a single change-point and selected the one with lowest DIC value, which places that change-point at the year 2018, three years after the change-point for elites. Again, the models appear to fit the data quite well, although arguably slightly worse than in the case of the Twitter networks (e.g., with AUC values for the best fitting change-point model computed at each one-year period above $0.840$ and overall AUC values above $0.890$). 

\begin{figure}[tbh]
\centering
\includegraphics[width=0.55\linewidth]{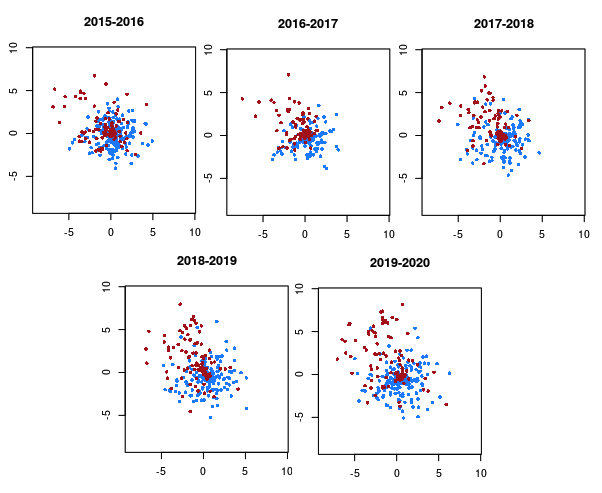}
\includegraphics[width=0.55\linewidth]{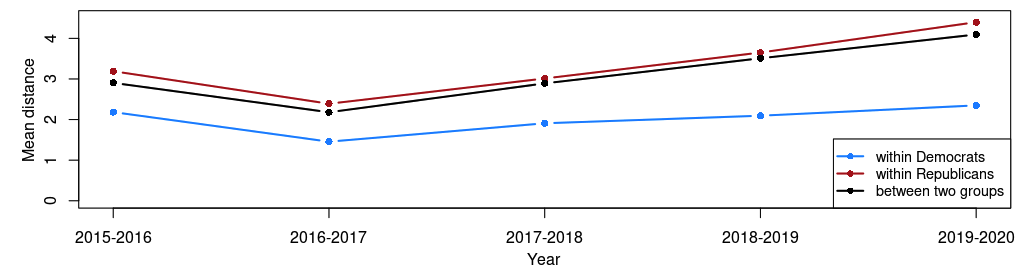}
\caption{\textit{Top}: Posterior means of latent positions for Reddit comment networks. Blue-Democrats, Red-Republicans. The dynamics of the latent positions is consistent with the evolution of within/between party densities seen in Figure \ref{reddit_den}. \textit{Bottom}: Mean latent distances within each group and between groups.}
\label{fig_latent_reddit}
\end{figure}

\begin{figure}[tbh!]
\centering
\includegraphics[width=0.55\linewidth]{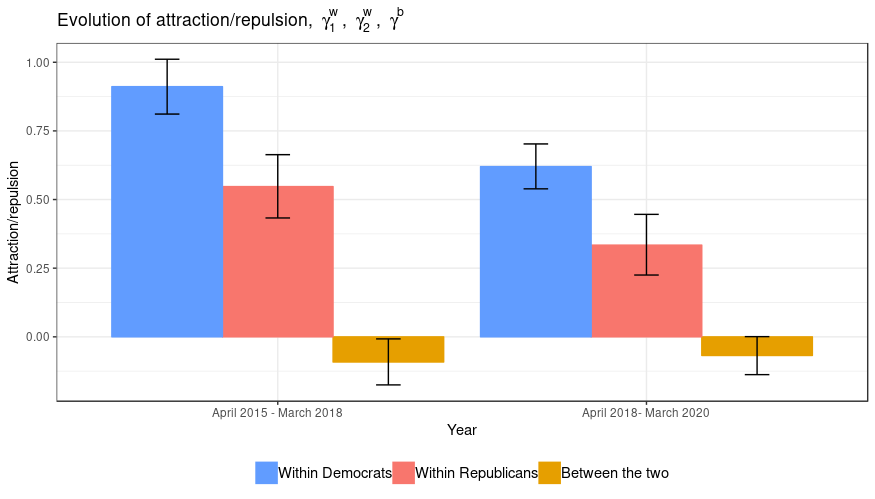}
\caption{Evolution of posterior means and 95\% CI for within-group attraction $\gamma_1^w$ for Republicans, $\gamma_2^w$ for Democrats, and between-group attraction/repulsion $\gamma^b$ in Reddit comment networks. Polarization across the two sets of Reddit users appeared for both two time periods, and the within-group attraction declined over time for both two groups. }
\label{fig_reddit_gamma}
\end{figure}

Table \ref{tab_para_reddit} and Figure \ref{fig_latent_reddit} show the results from fitting without a change-point, analogous to Table \ref{tab_para_twitter} and Figure \ref{fig_latent_twitter} for the Twitter data. Figure \ref{fig_reddit_gamma} displays the evolution of the with-in group coefficients for the two groups and the between-group coefficient, in analogy to Figure \ref{fig_twitter_gamma} for the Twitter data.  (Similarly, Figure 2 in the SI Appendix displays the evolution of edge persistence, in analogy to
Figure 1 in the SI Appendix.)

\begin{table}
\caption{Summary statistics for the posterior distribution of parameters using the whole sequence of Reddit networks from 2015 to 2020.}
\begin{threeparttable}
\centering
\begin{tabular}{*{7}{c}}
&$\hat \tau^2$ & $\hat \alpha$ & $\hat \delta$ & $\hat \gamma_1^w$ & $\hat \gamma_2^w$ & $\hat \gamma^b$  \\ 
\hline
Mean & 2.521& 3.079 & 0.937 & 0.748 & 0.401 & -0.128 \\
  SD &0.132& 0.011 & 0.010 & 0.032 & 0.041 & 0.027 \\ 
2.5\% Quantile &2.273& 3.056 & 0.917 & 0.686 & 0.321 & -0.182 \\ 
97.5\% Quantile &2.791& 3.101 & 0.957 & 0.811 & 0.482 & -0.075 \\ 
\hline
\end{tabular}
\begin{tablenotes}
\item 1-Democrats, 2-Republicans. The estimated variance $\hat \tau^2$ for the initial distribution of latent positions is much smaller than that for the Twitter data due to the initial Reddit network being much denser. This results indicate presence of edge persistence ($\hat \delta > 0$), higher within-group attraction for Democrats than Republicans ($\hat \gamma_1^w > \hat \gamma_2^w$ (mean difference = .347, SE = .035)), presence of between-group repulsion ($\hat \gamma^b <0$), and less magnitude of between-group repulsion than within-group attraction for both Democratic and Republican Reddit users ($|\hat \gamma^b| < |\hat \gamma_1^w|$ (mean difference = -.620, SE = .021), $|\hat \gamma^b| < |\hat \gamma_2^w|$ (mean difference = -.273, SE = .026)). 
\end{tablenotes}
\end{threeparttable}
\label{tab_para_reddit}
\end{table}

Some conclusions regarding evolution across the two time periods: 1) edge persistence increased (mean increase = .165, SE = .020); 2) between-group repulsion was present, demonstrating polarization across the sets of Democratic and Republican users of Reddit, though with some evidence that such polarization was mitigated (mean decrease = .023, SE = .054, P(decrease$>$0)=0.668) in the second time period starting in 2018;
3) while the two groups have moved away from one another, both experienced positive partisanship (within-group attraction) and became less concentrated over time, as both groups experienced a decline (Democrats: mean decrease = .291, SE = .066; Republicans: mean decrease = .213, SE = .082) in within-group attraction in the second time period; and 4) positive partisanship dominated the entire time period for both Democrat and Republican users of Reddit. 
The latter finding is particularly notable, since it suggests different polarization trends among the public than what we found above among members of Congress. Though they are different social media platforms and we have a shorter time-span on Reddit, the consistent dominance of positive partisanship for both Republicans and Democrats among the public and the dominance of negative partisanship among Republican elites over the same period suggests a disconnect between elites and the public, an early focus of debate in the polarization literature \citep{fiorina2005culture, abramowitz2008polarization}. 


\section{Discussion}

We develop a two-group coevolving latent space network with attractors (CLSNA) model for characterizing the dynamics of polarization in social networks. This model incorporates the effects of both attraction and repulsion by specifying appropriate attractor functions to explain the factors driving network interactions. It may be viewed as a type of causal modeling framework (in the spirit of, say, Granger causal modeling (\cite{granger1969investigating}), specifically designed to combine dynamical systems from mathematical modeling with principles of hierarchical statistical modeling. The former allows us to incorporate precise notions of attraction/repulsion relevant to polarization, while the latter permits principled and computationally tractable inferences in the form of statistical estimation, testing and prediction. 

While we focus on the context of polarization with the two-group version of CLSNA model, our proposed class of CLSNA models is a flexible framework which may incorporate a variety of attractor functions, making it general and quite broadly applicable to other co-evolutionary social dynamics where behaviors and beliefs impact social interactions, and vice versa. One limitation of our model, as implemented here, is that we assume the node set is fixed over time, which restricts our focus  on individuals who are active for the entire period of study. Those who come and go and stay active for only a certain period of time, which is common in practice, is not currently accounted for in our model. It is an interesting subject for future research to design dynamic network models allowing for varying set of nodes. 

Considering that the current model is defined for binary edges, and there are many other ways of constructing binary networks from Twitter and Reddit data besides the way we choose, it is of interest to assess the robustness of our results to different network constructions. We explore robustness by fitting models to sequences of networks constructed differently, for example, Twitter networks constructed from a static threshold with value 10, and Reddit networks constructed from using yearly averages as thresholds. Details on the robustness analysis are provided in Section 8 of the SI Appendix. In general the qualitative conclusions under the whole sequence analysis are quite robust except for two aspects:  the qualitative conclusion we have on the dominance of between-group repulsion for Republicans in Twitter and that on the presence of polarization in Reddit seem to be sensitive to the choice of network constructions. This speaks to the other limitation of our model in that it only accounts for binary networks, and some results could vary upon different dichotomizations. It is possible to extend the current model to weighted networks by using a link function to model the mean of the edge weights \citep{sewell2016latent}. However, this is for future research and beyond the scope of the present work. 

Model selection is an important step in the model fitting process, which pertains to choosing the number of dimensions of the latent space and finding the best place for the change-point. In this paper, we choose to use DIC for comparing fitted models due to its simplicity and popularity in the Bayesian community. However, there are a variety of other approaches for model selection, for example, using other information criteria (e.g., AIC, BIC, WAIC), and cross-validation. There is also work done, for example, \cite{vehtari2017practical}, on accounting for uncertainty in the information criteria. Such alternative criteria could of course be used in our work and we leave the details of the adaptation of such methods to the class of models studied in this paper as future work.

We also note that the networks studied in this paper are not sparse. While this is not a big issue for moderate size latent space models, it would potentially become an issue for other approaches such as Exponential Random Graph Models (ERGMs). It would certainly be of interest to incorporate sparseness in our models and study its effect. This is another topic for future research.

In this article, we focus the application of our model on two online longitudinal social networks, one of the political elite via Twitter, and one of the public, via Reddit. Our model has captured, disentangled and quantified two key aspects of polarization, positive and negative partisanship, as well as a concept in social network theory, edge persistence. Our results show that for members of Congress active on Twitter polarization across the two parties appeared throughout the past decade. We also study the time-varying aspects of attraction, repulsion and edge persistence by fitting a change-point version of our model. Results show that for Republican members of Congress, disengagement with the other party began to dominate engagement with their own in 2015, while disengagement with their own party also decreased at this time. Thus, among Republican members of Congress we find that increasing disengagement with the out-group party is not necessarily accompanied by strong in-group attachments. In fact, within-party forces for Republican members of Congress became negative after 2015. In contrast, for Democratic members of Congress positive partisanship was strongest throughout the entire period of study. We also find evidence of polarization among the public on Reddit. However, here positive partisanship dominated the full length of study for both Democrats and Republicans. Thus, the results provide only limited support for the increasing role of negative partisanship in polarization. In all, through the modeling and analyses of social media data of both the public and the political elite in the US, this work provides new insights into the nature and presence of polarization, as well as how positive and negative partisanship play roles and evolve over time.  

\section*{Acknowledgement}
We are very grateful to Luis Carvalho and Herbert Weisberg for their valuable insights and discussions. This research was supported by ARO award W911NF1810237, NSF DMS-2107856, and NSF SES-2120115.

\bibliographystyle{rss}
\bibliography{references, cantay_bib}
\end{document}


\maketitle
\sloppy

This supplement contains eight sections. 
In Section \ref{sec:0}, we provide a discussion on our choice of using hashtags in Twitter to define network interactions. In Section \ref{sec:1}, we present the details of the Bayesian method and the Markov chain Monte Carlo (MCMC) algorithm used to estimate the parameters. In Section \ref{sec:2-new}, we provide additional remarks on the rationale behind fixing $\sigma^2=1$. In Section \ref{sec:2}, we evaluate the performance of our proposed posterior-based inference approach through simulation and display the results. In Section \ref{sec:3sim-dic}, we evaluate the performance of our proposed parameter estimation procedure under the change-point setting where parameter values are time-varying. In Section \ref{sec:4}, we provide two figures mentioned in the main text, for the evolution of edge persistence in Twitter and Reddit data, respectively. 
In section \ref{sec:5}, we provide supplementary results in Twitter/Reddit data analysis, which are DIC values for model selection and a select handful of alternative analysis results for both Twitter and Reddit data with different choices of change-point. 
In section \ref{sec:6}, we provide a robustness check on the model results to network constructions.

\section{Polarization \&~Drawing Edges in Social Media Networks}
\label{sec:0}

We note in the manuscript and expand here that we recognize that the choice of edge construction can be consequential in studies of polarization. While the approaches vary, for the Twitter analysis we follow the literature that has taken advantage of hashtag sharing \citep{borge-holthoefer_content_2015, magdy__2016, weber_secular_2013, darwish_quantifying_2019, bovet_validation_2018, kusen_politics_2018}. 

We note first that we are not alone in making use of Twitter hashtags in the substantive context of polarization. Political scientists, among others researching social science topics, have used the hashtag feature of Twitter to explain different ways polarization manifests itself in networked online discourses 
\citep{garimella_polarization_2018, karlsen_echo_2017, abramowitz_is_2008, tucker_social_2018, ladd_why_2011, barbera_tweeting_2015}, and to quantify polarization in various American and international contexts 
\citep{borge-holthoefer_content_2015, magdy__2016, weber_secular_2013, darwish_quantifying_2019, masroor_polarization_2019, bovet_validation_2018, sharma_black_2013, zappavigna_discourse_2012, kusen_politics_2018, lai_debate_2015}. But the network construction approaches are varied. Researchers of polarization have used connections \citep{barbera_tweeting_2015, borge-holthoefer_content_2015}, mentions \citep{magdy__2016, conover_political_2011}, replies \citep{magdy__2016}, retweets \citep{garimella_polarization_2018, magdy__2016, weber_secular_2013, darwish_quantifying_2019, conover_political_2011}, and hashtags \citep{borge-holthoefer_content_2015, magdy__2016, weber_secular_2013, darwish_quantifying_2019, bovet_validation_2018, kusen_politics_2018} to construct networks in order to understand and quantify polarization in online discourses. 

We have intentionally chosen a data-rich and substantively meaningful measure of polarization in this context with the hashtag edge construction. Relative to direct mentions and retweets, hashtag sharing is more common, which provides greater evidence of engagement in the data. Relatedly, it is a socially easier activity, in so far as you do not have to directly address a fellow member, just the topic of interest. In this way it mirrors our measure of edges in the Reddit data, since those are constructed by sharing conversation threads and not necessarily direct interactions or replies.  It also provides a more equal playing field for junior and marginal members of Congress who may not feel comfortable directly engaging or calling-out more senior members, but would like to weigh-in on various issues. 

Another advantage of our approach is that we use both the entire population of Congresspeople active on Twitter in this period, as well as the entire population of hashtags used by them, instead of samples of users \citep{tyagi_affective_2020, chotisarn_bubble_2021, ozerim_discussing_2021, stewart_examining_2018} or samples of hashtags \cite{sharma_black_2013, lai_debate_2015, himelboim_birds_2013, kusen_politics_2018, lansdall-welfare_change-point_2016-1}. 

It is of perhaps most importance to note that the indirect hashtag approach here operationalizes polarization in terms of discontinued topic engagement. Specifically, polarization (as evinced here) is when members of opposite parties discontinue to engage the same topic, which is an arguably higher bar than when they discontinue direct mentions of an individual. While legislators may stop talking to one another at times, one can think of few greater condemnations of a democratic system than refusing to engage the same topics.  

More generally, we note that the importance and nature of indirect engagements (like hashtag sharing) is widely studied, including in the context of conversations on Twitter \cite[e.g.,][]{bruns_structural_2014, naaman_hip_2011, wilkinson_trending_2012, giglietto_hashtag_2017, bode_candidate_2015}, as well as in online and in-person engagements \cite[e.g.,][]{dyck_know_2014, garner_ambivalence_2013, dovidio_social_2017}.  

Finally, we note here that we remain agnostic on the causal mechanism of partisan polarization in our analysis of social media network topic engagement. Whether the particular online interactions we observe are predominantly affective or policy driven are beyond the scope of this work, but a worthy topic of interest for future research.


In sum, we recognize that alternative measures of edges may lead to different conceptions of polarization, which may arrive at different results \citep[see, e.g.,][]{Lelkes2016}. However, we believe our edge constructions are well justified, amply used in the literature, and provide important substantive insights into partisan polarization in social media interactions.

\section{Markov Chain Monte Carlo Estimation}\label{sec:1}
Given an observed network time series $\{G_t\}_{t=1}^T$ or, equivalently, a time series of the corresponding adjacency matrices $\{Y_t\}_{t=1}^T$, we wish to make inference of the latent positions $\bm Z_t$ and model parameters $\bm \theta = (\alpha, \delta, \gamma^w_1, \gamma^w_2, \gamma^b, \tau^2, \sigma^2)$. Given the hierarchical nature of our model, it is natural to adopt a Bayesian approach. In the meanwhile, estimation within Bayesian framework can address the issues of lack of identifiability in the likelihood function of our model. Specifically, in our model, the probability of observing $Y_{1:T}$ conditional on all unknowns is 
\begin{equation}
\begin{split}
P(Y_{1:T} \,|\, \bm Z_{1:T}, \alpha, \delta)
& = P(Y_1 \,|\, \bm Z_1) \prod_{t=2}^TP(Y_t \,|\, \bm Z_t, Y_{t-1}) \\
&= 
\prod_{i \neq j}P_{\alpha}(Y_{1, ij} \,|\, \bm Z_{1,i}, \bm Z_{1, j}) \cdot 
\prod_{t=2}^T \prod_{i \neq j}   P(Y_{t,ij} \,|\, \bm Z_{t, i}, \bm Z_{t, j}, Y_{t-1, ij})\\
& =  \prod_{i \neq j} \left( P_{\alpha}(Y_{1, ij} \,|\, \bm Z_{1,i}, \bm Z_{1, j}) 
\prod_{t=2}^T  P_{\alpha, \delta}(Y_{t,ij} \,|\, \bm Z_{t, i}, \bm Z_{t, j}, Y_{t-1, ij}) \right).
\end{split}
\label{eq:1}
\end{equation}
Notice that  
\begin{equation}
\begin{split}
& P(Y_{t,ij} \,|\, \bm Z_{t, i}= z_{t, i}, \bm Z_{t, j}= z_{t, j}, Y_{t-1, ij}, \alpha, \delta) \\
& = \left(\frac{\exp(\eta_{t,ij})}{1+\exp(\eta_{t,ij})} \right)^{Y_{t, ij}}
\left(\frac{1}{1+\exp(\eta_{t,ij})}\right)^{1-Y_{t, ij}} \\
& = \frac{\exp(Y_{t,ij}\eta_{t,ij})}{1+\exp(\eta_{t,ij})},
\end{split}
\end{equation}
where 
\begin{equation}
\begin{split}
    \eta_{t, ij} := \text{logit}(p_{t, ij}) &= \log \left(
    \frac{P(Y_{t, ij}=1 | \bm Z_t, Y_{t-1}, \bm \theta)}{P(Y_{t, ij}=0| \bm Z_t, Y_{t-1}, \bm \theta)} \right) \\
    &=
\begin{cases}
\alpha + \delta Y_{t-1,ij} - s(\bm z_{t,i}, \bm z_{t,j})  & \text{ if } t > 1\\
\alpha - s(\bm z_{t,i}, \bm z_{t,j}) & \text{ if } t = 1.
\end{cases}
\end{split}
\end{equation}

Note that if $\exists$ a set of distances and $\alpha$, $\delta$ s.t. $\eta_{t, ij} >0 $ when $Y_{t,ij}=1$ and $\eta_{t, ij} <0 $ when $Y_{t,ij}=0$, then by rescaling, i.e. let $\tilde z_t = \kappa z_t$, and $\tilde \alpha = \kappa \alpha$, $\tilde \delta = \kappa \delta$, $\kappa > 0$, the conditional distribution of $Y_{t,ij}$ becomes
 \begin{equation}
\begin{split}
 P(Y_{t,ij} \,|\, \bm Z_{t, i}=\tilde z_{t, i}, \bm Z_{t, j}=\tilde z_{t, j}, Y_{t-1, ij}, \tilde \alpha, \tilde \delta)  = \frac{\exp(Y_{t,ij}\kappa \eta_{t,ij})}{1+\exp(\kappa\eta_{t,ij})} \rightarrow 1, \kappa \rightarrow \infty.
\end{split}
\end{equation}
Therefore, the likelihood \eqref{eq:1} is not identifiable, as $z_t$, $\alpha$ and $\delta$ can be scaled arbitrarily. However, the posterior distribution $\pi(\bm Z_{1:T}, \bm \theta \,|\, Y_{1:T})$ is identifiable by imposing appropriate prior distributions on the unknown parameters, which can be viewed as a set of constrains to prevent parameters from scaling arbitrarily. To sample from the posterior distribution, we implement an adaptive Metropolis-Hastings (MH) within Gibbs MCMC scheme, where samples are drawn from the full conditional distributions iteratively. These conditional distributions are either known in closed form or up to a normalizing constant. We sample directly if the conditional distribution is known, otherwise we sample via MH using an adaptive normal random walk proposal. We provide details in the following subsections, where two issues are addressed, one regarding the lack of identifiability for the conditional distribution of $\gamma_1^w$ and $\gamma_2^w$ in connection with the scaling of $\sigma^2$ and latent positions $\bm Z_t$, the other one regarding rotational invariance in the latent space. 

\subsection{Posterior and Full Conditional Distributions}
The posterior distribution of latent positions and parameters is 
\begin{equation}
\begin{split}
& \pi(\bm Z_{1:T}, \bm \theta \,|\, Y_{1:T}) \\
& \propto P(Y_{1:T}, \bm Z_{1:T} \,|\, \bm \theta) \pi(\bm \theta) \\
& \propto 
\left( 
\prod_{t=2}^T P(\bm Z_t \,|\, \bm Z_{t-1}, Y_{t-1}) P(Y_t \,|\, \bm Z_t, Y_{t-1}) 
\right)
P(Y_1 \,|\, \bm Z_1)P(\bm Z_1) \pi(\bm \theta) \\
& = 
\left(
\prod_{t=2}^T 
\prod_{i=1}^N \left( P(\bm Z_{t,i} \,|\, \bm Z_{t-1}, Y_{t-1}) \prod_{j: j \neq i} P(Y_{t,ij} \,|\, \bm Z_{t, i}, \bm Z_{t, j}, Y_{t-1, ij}) \right)
\right) \cdot\\
& \qquad \qquad \qquad \qquad \qquad \qquad \qquad \qquad \qquad \qquad \qquad \qquad 
P(Y_1 \,|\, \bm Z_1)P(\bm Z_1) \pi(\bm \theta) \\
&=
\prod_{t=2}^T 
\prod_{i=1}^N P_{\gamma^w_{\pi(i)}, \gamma^b, \sigma^2}(\bm Z_{t,i} \,|\, \bm Z_{t-1}, Y_{t-1}) 
\cdot
\prod_{t=2}^T 
\prod_{i\neq j}
P_{\alpha, \delta}(Y_{t,ij} \,|\, \bm Z_{t, i}, \bm Z_{t, j}, Y_{t-1, ij})
\cdot  \\
& 
\qquad \qquad \qquad \qquad \qquad \qquad \qquad \qquad 
\prod_{i \neq j}P_{\alpha}(Y_{1, ij} \,|\, \bm Z_{1,i}, \bm Z_{1, j})
\cdot
\prod_{i=1}^N P_{\tau^2}(\bm Z_{1,i}) 
\cdot
\pi(\bm \theta)
\end{split}
\end{equation}

We set the priors on the parameters as follows: assume that 
$\alpha \sim \hbox{N}(\nu_{\alpha}, \xi^2_{\alpha})$, $\delta \sim \hbox{N}(\nu_{\delta}, \xi^2_{\delta})$, 
$\gamma_1^w, \gamma_2^w  \sim \hbox{N}(\nu_{\gamma^w}, \xi^2_{\gamma^w} )$, 
$ \gamma^b \sim \hbox{N}(\nu_{\gamma^b}, \xi^2_{\gamma^b} )$ , $\tau^2 \sim \hbox{IG}(\eta_{\tau}, \phi_{\tau})$, and $\sigma^2 \sim \hbox{IG}(\eta_{\sigma}, \phi_{\sigma})$ 
where N(), IG() are normal and inverse gamma distributions, respectively. The full conditional distributions follow.

\subsubsection{Sampling $Z_{t, i}$} The conditional distribution for $\bm Z_{t, i}$ is 
\begin{equation}
\begin{split}
&\pi(\bm Z_{t, i} \,|\, Y_{1:T}, \bm \theta, \bm Z_{\setminus t, i}) \propto	\\
&	\begin{cases}
	\left( \prod_{j:j\neq i}s_{t, ij}s_{t, ji}\right) 
	\cdot N(\bm Z_{t+1, i} \, |\,  \bm Z_{t,i} + \gamma^w_{\pi(i)} A_i^w(\bm Z_{t},Y_{t}) + \gamma^b A_i^b(\bm Z_{t},Y_{t}), \sigma^2 I_p) \cdot   \\
    \qquad 
    \prod_{j\in \mathcal{N}_{i, t} }N(\bm Z_{t+1, j} \, |\,  \bm Z_{t,j} + \gamma^w_{\pi(j)} A_j^w(\bm Z_{t},Y_{t}) + \gamma^b A_j^b(\bm Z_{t},Y_{t}), \sigma^2 I_p) \cdot \\
    \qquad \qquad \qquad\qquad \qquad \qquad\qquad \qquad \qquad
   N(\bm Z_{t,i} \,|\, \bm 0, \tau^2 I_p) 
\qquad\qquad\qquad \text{if } t=1 \\
	\left( \prod_{j:j\neq i}s_{t, ij}s_{t, ji}\right) 
	\cdot N(\bm Z_{t+1, i} \, |\,  \bm Z_{t,i} + \gamma^w_{\pi(i)} A_i^w(\bm Z_{t},Y_{t}) + \gamma^b A_i^b(\bm Z_{t},Y_{t}), \sigma^2 I_p)   \cdot\\
	\qquad 
	\prod_{j\in \mathcal{N}_{i, t} }N(\bm Z_{t+1, j} \, |\,  \bm Z_{t,j} + \gamma^w_{\pi(j)} A_j^w(\bm Z_{t},Y_{t}) + \gamma^b A_j^b(\bm Z_{t},Y_{t}), \sigma^2 I_p) \cdot \\
	\qquad 
	N(\bm Z_{t, i} \, |\,  \bm Z_{t-1,i} + \gamma^w_{\pi(i)} A_i^w(\bm Z_{t-1},Y_{t-1}) + \gamma^b A_i^b(\bm Z_{t-1},Y_{t-1}), \sigma^2 I_p)
\quad \text{ if } 1<t<T \\
	\left( \prod_{j:j\neq i}s_{t, ij}s_{t, ji}\right) 
	\cdot 
	N(\bm Z_{t, i} \, |\,  \bm Z_{t-1,i} + \gamma^w_{\pi(i)} A_i^w(\bm Z_{t-1},Y_{t-1}) + \\
	\qquad 	\qquad 	\qquad 	\qquad 	\qquad 	\qquad 
	\qquad 	\qquad 	\qquad 
	\gamma^b A_i^b(\bm Z_{t-1},Y_{t-1}), \sigma^2 I_p)  \qquad \text{ if } t = T
	\end{cases}
\end{split}
\end{equation}
where $s_{t,ij} := P(Y_{t,ij} \,|\, \bm Z_{t, i}= z_{t, i}, \bm Z_{t, j}= z_{t, j}, Y_{t-1, ij}, \alpha, \delta) = \frac{\exp(Y_{t,ij}\eta_{t,ij})}{1+\exp(\eta_{t,ij})}$, and $\mathcal{N}_{i, t}$ is the set of neighbors of node $i$ at time $t$.

\subsubsection{Sampling $\alpha$ and $\delta$}

The conditional distribution for $\alpha$ is 
\begin{equation}
\begin{split}
\pi(\alpha\,|\, Y_{1:T}, \bm Z_{1:T}, \bm \theta_{\setminus \alpha}) 
&\propto \left( \prod_{t=1}^T \prod_{i\neq j} s_{t, ij} \right) \cdot \pi(\alpha) \\
& = \left( \prod_{t=1}^T \prod_{i\neq j} s_{t, ij} \right) \cdot N(\alpha \,|\, \nu_{\alpha}, \xi_{\alpha}^2)
\end{split}
\end{equation}

The conditional distribution for $\delta$ is 
\begin{equation}
\begin{split}
\pi(\delta\,|\, Y_{1:T}, \bm Z_{1:T}, \bm \theta_{\setminus \delta}) 
& \propto \left( \prod_{t=2}^T \prod_{i\neq j} s_{t, ij} \right) \cdot \pi(\delta) \\
& = \left( \prod_{t=2}^T \prod_{i\neq j} s_{t, ij} \right)   \cdot N(\delta \,|\, \nu_{\delta}, \xi_{\delta}^2)
\end{split}
\end{equation}

\subsubsection{Sampling $\gamma^w_1$, $\gamma^w_2$} \label{sec:1-1-3}
Note that for the conditional distribution of $\gamma^w_1$, we have  
 \begin{equation}
\begin{split}
&\pi(\gamma^w_1\,|\, Y_{1:T}, \bm Z_{1:T}, \bm \theta_{\setminus \gamma^w_1}) \\
&\propto
\prod_{t=2}^T 
\prod_{i=1}^N P_{\gamma^w_{\pi(i)},\gamma^b, \sigma^2}(\bm Z_{t,i} \,|\, \bm Z_{t-1}, Y_{t-1}) \cdot \pi(\gamma^w_1) \\
& \propto
\prod_{t=2}^T 
\prod_{i: \pi(i)=1}
N(\bm Z_{t, i} \, |\,  \bm Z_{t-1,i} + \gamma^w_1 A_i^w(\bm Z_{t-1},Y_{t-1} + \\
& \qquad\qquad\qquad\qquad\qquad\qquad
\gamma^b A_i^b(\bm Z_{t-1},Y_{t-1})), \sigma^2 I_p)  
\cdot
N(\gamma^w_1 \,|\,\nu_{\gamma^w}, \xi_{\gamma^w}^2),
\end{split}
\end{equation}
and from here we can get a closed form distribution. Similarly for  $\gamma^w_2$.

Hence the conditional distributions for $\gamma^w_1$, $\gamma^w_2$ are
\begin{equation}
\begin{split}
& \gamma^w_1\,|\, Y_{1:T}, \bm Z_{1:T}, \bm \theta_{\setminus \gamma^w_1} \sim \\
& N
\left(
\frac{
\frac{1}{\sigma^2}
\sum\limits_{t=2}^T \sum\limits_{i:\pi(i)=1} \bm a_{t,i}^T \bm b_{t,i}
+ \frac{1}{\xi_{\gamma^w}^2} \nu_{\gamma^w}
}
{
\frac{1}{\sigma^2}
\sum\limits_{t=2}^T \sum\limits_{i:\pi(i)=1} \bm b_{t,i}^T \bm b_{t,i}
+ \frac{1}{\xi_{\gamma^w}^2}
}, 
\frac{1}
{
(\frac{1}{\sigma^2}
\sum\limits_{t=2}^T \sum\limits_{i:\pi(i)=1}\bm b_{t,i}^T \bm b_{t,i}
+ \frac{1}{\xi_{\gamma^w}^2})^{1/2}
}
\right)
\end{split}
\end{equation}

\begin{equation}
\begin{split}
& \gamma^w_2\,|\, Y_{1:T}, \bm Z_{1:T}, \bm \theta_{\setminus \gamma^w_1} \sim \\
& N
\left(
\frac{
\frac{1}{\sigma^2}
\sum\limits_{t=2}^T \sum\limits_{i:\pi(i)=2} \bm a_{t,i}^T \bm b_{t,i}
+ \frac{1}{\xi_{\gamma^w}^2} \nu_{\gamma^w}
}
{
\frac{1}{\sigma^2}
\sum\limits_{t=2}^T \sum\limits_{i:\pi(i)=2} \bm b_{t,i}^T \bm b_{t,i}
+ \frac{1}{\xi_{\gamma^w}^2}
}, 
\frac{1}
{
(\frac{1}{\sigma^2}
\sum\limits_{t=2}^T \sum\limits_{i:\pi(i)=2}\bm b_{t,i}^T \bm b_{t,i}
+ \frac{1}{\xi_{\gamma^w}^2})^{1/2}
}
\right)
\end{split}
\end{equation}
where $\bm a_{t,i} = z_{t,i}- z_{t-1,i}-\gamma^b A_i^b(z_{t-1}, Y_{t-1}) $, $\bm b_{t,i} =  A_i^w(z_{t-1}, Y_{t-1})$. 

Notice that if we rescale $\tilde z_t = \kappa z_t$, and $\tilde \sigma^2 = \kappa^2 \sigma^2$, we obtain the same conditional distributions of $\gamma^w_1$ and $\gamma^w_2$ since their means and variances remain unaltered. Although the joint posterior distribution is identifiable, the lack of identifiability in these two conditional distributions will cause trouble in the convergence of the Gibbs sampler. To address these forms of rescaling, we fix $\sigma^2=1$.

\subsubsection{Sampling $\gamma^b$}
 \begin{equation}
\begin{split}
\pi(\gamma^b\,|\, Y_{1:T}, \bm Z_{1:T}, \bm \theta_{\setminus \gamma^b}) 
&\propto
\prod_{t=2}^T 
\prod_{i=1}^N P_{\gamma^w_{\pi(i)},\gamma^b, \sigma^2}(\bm Z_{t,i} \,|\, \bm Z_{t-1}, Y_{t-1}) \cdot \pi(\gamma^b) \\
& \propto
\prod_{t=2}^T 
\prod_{i=1}^N
N(\bm Z_{t, i} \, |\,  \bm Z_{t-1,i} + \gamma^w_{\pi(i)} A_i^w(\bm Z_{t-1},Y_{t-1} + \\
& \qquad\qquad
\gamma^b A_i^b(\bm Z_{t-1},Y_{t-1})), \sigma^2 I_p)  
\cdot
N(\gamma^b \,|\,\nu_{\gamma^b}, \xi_{\gamma^b}^2)
\end{split}
\end{equation}

The conditional distribution for $\gamma^b$ is 
\begin{equation}
\gamma^b\,|\, Y_{1:T}, \bm Z_{1:T}, \bm \theta_{\setminus \gamma^b} \sim N
\left(
\frac{
\frac{1}{\sigma^2}
\sum\limits_{t=2}^T \sum\limits_{i=1}^N \bm c_{t,i}^T \bm d_{t,i}
+ \frac{1}{\xi_{\gamma^b}^2} \nu_{\gamma^b}
}
{
\frac{1}{\sigma^2}
\sum\limits_{t=2}^T \sum\limits_{i=1}^N \bm d_{t,i}^T \bm d_{t,i}
+ \frac{1}{\xi_{\gamma^b}^2}
}, 
\frac{1}
{
(\frac{1}{\sigma^2}
\sum\limits_{t=2}^T \sum\limits_{i=1}^N \bm d_{t,i}^T \bm d_{t,i}
+ \frac{1}{\xi_{\gamma^b}^2})^{1/2}
}
\right)
\end{equation}
where $\bm c_{t,i} = z_{t,i}- z_{t-1,i}-\gamma^w_{\pi(i)} A_i^w(z_{t-1}, Y_{t-1}) $, $\bm d_{t,i} =  A_i^b(z_{t-1}, Y_{t-1})$.

\subsubsection{Sampling $\tau^2$ and $\sigma^2$}
The conditional distributions for $\tau^2$ and $\sigma^2$ are
\begin{equation}
    \tau^2 \,|\, Y_{1:T}, \bm Z_{1:T}, \bm \theta_{\setminus \tau^2} 
 \sim 
IG(\eta_\tau + \frac{np}{2}, \,\phi_\tau + \frac{1}{2} \sum_{i=1}^N ||\bm Z_{1, i}||^2) 
\end{equation}

\begin{equation}
\begin{split}
  \sigma^2 \,|\, Y_{1:T}, \bm Z_{1:T}, \bm \theta_{\setminus \sigma^2} 
&\sim 
\quad IG(\eta_\sigma + \frac{np(T-1)}{2}, \phi_\sigma + \frac{1}{2} \sum_{t=2}^T\sum_{i=1}^N ||\bm Z_{t, i} - \\
& \qquad \qquad
\bm Z_{t-1, i} - \gamma^w A_i^w(\bm Z_{t-1}, Y_{t-1})
 - \gamma^b A_i^b(\bm Z_{t-1}, Y_{t-1})||^2)  
\end{split}
\end{equation}
Note that in our implementation of the posterior sampler, we set $\sigma^2=1$ to fix the scaling issues in $\gamma^w_1$, $\gamma^w_2$ sampling, hence no sample is drawn from the conditional distribution of $\sigma^2$.

\subsection{Rotational Invariance}
The last issue to address around posterior sampling is that the posterior distribution is invariant to rotations, reflections and translation of latent positions. Following \cite{sewell2015latent} and \cite{hoff2002latent}'s work, we perform a Procrustes transformation to reorient the sampled latent positions. 

Recall that $\bm Z_t \in R^{N\times p}$, where $N$ is the number of nodes in the network and $p$ is the dimension of the latent space. Let $\bm{\mathcal{Z}} = [\bm Z_1^T, \bm Z_2^T,..., \bm Z_T^T]^T \in R^{(nT) \times p}$, and $\bm{\mathcal{Z}}^{(k)}$ be the samples at the $k$-th iteration. In our posterior sampling, we do the following,
\begin{enumerate}
    \item Take $\bm{\mathcal{Z}^{(0)}}$ (centered) as reference positions. 
    \item For each $k>0$, perform Procrustes transformation on the new draws $\bm{\mathcal{Z}^{(k)}}$ (centered), i.e.,
$$\bm{\mathcal{Z}^{(k)}} \leftarrow \argmin_{\bm{\mathcal{Z}^\star}: \bm{\mathcal{Z}^\star}= \bm{\mathcal{Z}^{(k)}} R}
tr(\bm{\mathcal{Z}^{(0)}} - \bm{\mathcal{Z}^\star})^T (\bm{\mathcal{Z}^{(0)}} - \bm{\mathcal{Z}^\star}),$$
where $R$ is a rotation matrix, and $\bm{\mathcal{Z}^\star}$ is some rotation of $\bm{\mathcal{Z}^{(k)}}$. The transformed latent positions are a reorientation of the previous draws, which preserve the distance between any actors at any time and are the closest to the reference positions compared with other rotations. 
\end{enumerate}

\subsection{Posterior Sampling via Adaptive Metropolis-Hastings within Gibbs} Combining all the pieces discussed above, our adaptive Metropolis-Hastings within Gibbs sampling algorithm is as follows. Set initial values for $\bm Z_{1:T}$, $\alpha, \delta, \gamma_1^w, \gamma_2^w, \tau^2$, and fix $\sigma^2 = 1$. Then for every iteration:
\begin{enumerate}
\item For $t = 1,\cdots, T$ and for $i=1, \cdots N$, draw $Z_{t, i}$ via MH using an adaptive normal random walk proposal.
\item Draw $\alpha$ via MH using an adaptive normal random walk proposal.
\item Draw $\delta$ via MH using an adaptive normal random walk proposal.
\item Draw $\tau^2$ directly from its conditional inverse gamma distribution.
\item Draw $\gamma^w_1$ directly from its conditional normal distribution.
\item Draw $\gamma^w_2$ directly from its conditional normal distribution.
\item Draw $\gamma^b$ directly from its conditional normal distribution.
\end{enumerate}
The latent positions as well as $\alpha$, $\delta$ are updated using the adaptive normal random walk proposal \citep{rosenthal2011optimal}. As an example, for variable $\alpha$ being updated at iteration $k$, we propose a new value $\alpha^\prime$ drawn from normal distribution centered at $\alpha^{(k-1)}$, with standard deviation $\exp(s^{(k)})$ determined at previous iteration, and calculate the acceptance ratio $R^{(k)} = \frac{\pi(\alpha^\prime|\cdots)}{\pi(\alpha|\cdots)}$, then accept $\alpha^\prime$ with probability $\min\{1, R^{(k)}\}$. The notion of `adaptive' comes from the fact that the tuning parameter $s^{(k)}$ is updated at every iteration with $s^{(k)} = s^{(k-1)} + \frac{1}{(k-1)^{0.8}}\cdot \left[\min\left(1, R^{(k-1)}\right) - 0.234\right]$, which adjusts the scale of the random walk proposal based on the acceptance ratio, and prevents it from being too small causing the chain moving slowly or too large causing very high rejection rate. 

The pseudocode for the MCMC algorithm we proposed is provided in Algorithm \ref{al1}.


\begin{algorithm}
 \hspace*{\algorithmicindent} \textbf{Input}: Network time-series $Y_{1:T}$, total number of nodes $N$. Initial values $\bm \theta^{(0)}$, $\bm Z_{1:T}^{(0)}$ for parameters and latent positions. \\
 \hspace*{\algorithmicindent} Number of samples $N_{samples}$. Tuning parameters $s_1 \in R$, $s_2\in R$, and $\bm s\in R^{T \times N}$.\\
 \hspace*{\algorithmicindent} \textbf{Output}: $\bm \theta^{(1)}, \cdots, \bm \theta^{(N_{sample)}}$
\begin{algorithmic}[1]
    \State Set initial values for $\bm Z_{1:T}$, $\alpha, \delta, \gamma_1^w, \gamma_2^w, \tau^2$, and fix $\sigma^2 = 1$.
    \For{$k \leftarrow 1$ to $N_{sample}$}
        \State $\bm \theta^{(k)} \leftarrow \bm \theta^{(k-1)} $, $\bm Z_{1:T}^{(k)} \leftarrow \bm Z_{1:T}^{(k-1)}$
        \For{$t \leftarrow 1$ to $T$} 
            \For{$i \leftarrow 1$ to $N$}
                    \State Generate $Z_{t,i}^{\prime} = Z_{t,i}^{(k-1)} + N(0, \exp(s_{t,i}))$, $u\sim U(0, 1)$ \Comment{Get $k$-th sample for $Z_{t,i}$}
                    \State Calculate acceptance ratio $\log R = \log \pi(Z_{t,i}^\prime|\bm \theta^{(k)}, \bm Z_{1:T}^{(k)} \setminus Z_{t,i}^{(k)}) - \log \pi(Z_{t,i}^{(k-1)}|\bm \theta^{(k)}, \bm Z_{1:T}^{(k)}\setminus Z_{t,i}^{(k)})$
                \If{$\log u <\min(0, \log R)$}
                    \State $Z_{t,i}^{(k)} \leftarrow Z_{t,i}^\prime$
                    \Else
                    \State $Z_{t,i}^{(k)} \leftarrow Z_{t,i}^{(k-1)}$
                \EndIf
                \State $s_{t,i} \leftarrow s_{t,i} + \frac{1}{k^{0.8}}\cdot \left[\min\left(1, \exp(\log R)\right) - 0.234\right]$ \Comment{Update tuning parameter for $Z_{t,i}$}
            \EndFor
        \EndFor
        \State Generate $\alpha^{\prime} = \alpha^{(k-1)} + N(0, \exp(s_1))$, $u\sim U(0, 1)$ \Comment{Get $k$-th sample for $\alpha$}
        \State Calculate acceptance ratio $\log R_1 = \log \pi(\alpha^\prime|\bm \theta^{(k)}\setminus \alpha^{(k)}, \bm Z_{1:T}^{(k)}) - \log \pi(\alpha^{(k-1)}|\bm \theta^{(k)}\setminus \alpha^{(k)}, \bm Z_{1:T}^{(k)})$
        \If{$\log u <\min(0, \log R_1)$}
            \State $\alpha^{(k)} \leftarrow \alpha^\prime$
        \Else
            \State $\alpha^{(k)} \leftarrow \alpha^{(k-1)}$
        \EndIf
        \State $s_1 \leftarrow s_1 + \frac{1}{k^{0.8}}\cdot \left[\min\left(1, \exp(\log R_1)\right) - 0.234\right]$ \Comment{Update tuning parameter for $\alpha$}
        \State Generate $\delta^{\prime} = \delta^{(k-1)} + N(0, \exp(s_2))$, $u\sim U(0, 1)$ \Comment{Get $k$-th sample for $\delta$}
        \State Calculate acceptance ratio $\log R_2 = \log \pi(\delta^\prime|\bm \theta^{(k)}\setminus \delta^{(k)}, \bm Z_{1:T}^{(k)}) - \log \pi(\delta^{(k-1)}|\bm \theta^{(k)}\setminus \delta^{(k)}, \bm Z_{1:T}^{(k)})$
        \If{$\log u <\min(0, \log R_2)$}
            \State $\delta^{(k)} \leftarrow \delta^\prime$
        \Else
            \State $\delta^{(k)} \leftarrow \delta^{(k-1)}$
        \EndIf
        \State $s_2 \leftarrow s_2 + \frac{1}{k^{0.8}}\cdot \left[\min\left(1, \exp(\log R_2)\right) - 0.234\right]$ \Comment{Update tuning parameter for $\delta$}
        \State Obtain the $k$-th samples for $\tau^2$, $\gamma^w_1$, $\gamma^w_2$ and $\gamma^b$ by sampling directly from its conditional distribution using the most updated parameter values.
    \EndFor
\end{algorithmic}
\caption{\textsc{Adaptive Metropolis-Hastings within Gibbs Sampler}}
\label{al1}
\end{algorithm}

\subsection{MCMC settings} 
The priors on $\alpha$ and $\delta$ were chosen to be $N(0, 100)$ to keep it flat and uninformative. We chose the priors on $\gamma_1^w, \gamma_2^w$ to be $N(0.5, 100)$ and $\gamma^b$ to be $N(-0.5, 100)$, to reflect the prior belief of polarization, however these are also quite uninformative given the large variance. The prior on $\tau$ was chosen to be $IG(2.05, 1.05\sum_{i=1}^N ||\bm Z^{(1)}_{1,i}||^2/(Np))$ to be flat and uninformative following \cite{sewell2015latent}'s suggestion.

For the initialization, $\alpha, \delta, \gamma_1^w, \gamma_2^w, \gamma^b$ were initialized to be $0, 0, 0.5, 0.5, -0.5$, respectively. The estimation is quite robust to the initialization of these parameters, in our experience. All latent positions $\bm Z^{(1)}_{1:T}$ are initialized using the generalized multidimensional scaling (GMDS) method proposed by 
\cite{sarkar2005dynamic}, which first initializes $\bm Z^{(1)}_{1}$ through classical multidimensional scaling with similarity matrix being the shortest paths in $Y_1$, then initializes $\bm Z^{(1)}_{t}, t>1$ to be consistent with $Y_t$ and have similar pairwise distances as $\bm Z^{(1)}_{t-1}$. $\tau$ was initialized to be $\frac{1}{Np}\sum_{i=1}^N ||\bm Z^{(1)}_{1,i}||^2$ using the initial latent positions $\bm Z^{(1)}_{1}$.

For the results in the article, we set the number of MCMC iterations to be $50,000$ and removed a burn-in of $15,000$ samples. We set the initial value of the tuning parameter $s^{(1)}$ for $\alpha$ and $\delta$ to be $2$, and for $\bm Z_{1:T}$ to be $4$.

\section{Additional Remarks on the Rationale Behind Fixing $\sigma^2=1$}\label{sec:2-new}

The purpose of fixing $\sigma^2=1$ is to solve the issue of a lack of identifiability appearing in the model in connection with the scaling of $\sigma^2$, latent positions and parameters. Besides being a solution to a technical problem in MCMC as mentioned in Section \ref{sec:1-1-3}, it essentially enforces that the model, specifically $\hbox{logit}(p_{t,ij})$, be scale-invariant w.r.t. $\sigma^2$. That is, let $z_{t}, \alpha, \delta, \hbox{logit}(p_{t,ij})$ be those defined in equations (1)-(3) in the main text, and define 
\begin{equation}
   \tilde {\bm z_t} := \bm {z_t}/\sigma, \tilde\alpha := \alpha/\sigma, \tilde\delta := \delta / \sigma, 
\end{equation}
and $\tilde p_{t,ij}$ to be the probability of an edge under the parameterization $\bm{\tilde z_t}$, $\tilde \alpha$, $\tilde \delta$.
Provided that $\bm z_t | \bm z_{t-1} \sim N(\mu(\bm z_{t-1}, ...), \sigma^2I_p)$, where $\mu(\cdot)$ is the mean vector defined in equation (3) in the main text, we have 
\begin{equation}
   \bm{\tilde z_t }| \bm{\tilde z_{t-1}} \sim N(\mu(\bm {\tilde z_{t-1}}, ...), I_p),
\end{equation}
so given that $\bm z_t$ corresponds to the model with arbitrary $\sigma^2$, $\bm{\tilde{z_t}}$ corresponds to the model with $\sigma^2 = 1$.
We can also derive that 
\begin{equation}
\begin{split}
\hbox{logit}(p_{t,ij}) 
   & =  \alpha + \delta Y_{t-1,ij} - s(\bm z_{t,i}, \bm z_{t,j}) \\
   & = {\alpha} + {\delta} Y_{t-1,ij} -  \sigma s(\bm{\tilde z_{t,i}}, \bm{\tilde z_{t,j}})  \\
   & = \sigma (\tilde \alpha +\tilde{\delta} Y_{t-1,ij} - s(\bm{\tilde z_{t,i}}, \bm{\tilde z_{t,j}}) ) \\
   &= \sigma \hbox{logit}(\tilde p_{t,ij}).
\end{split}
\end{equation}
Hence, by rescaling latent positions and the variance, the probabilities of an edge differ by a multiplicative factor on the logit-scale, which could lead to an identifiability issue as $\hbox{logit}(p_{t,ij})$ can become arbitrarily large or small under the effect of the volatility $\sigma$. Fixing $\sigma^2=1$ will make the model scale-free, thus identifiable. The fixed value here is also related to the coefficient of the similarity function $s(\cdot, \cdot)$ being equal to $-1$. We choose the fixed value to be $1$ to make sure the coefficient of $s(\bm{\tilde{z}_{t,i}},\bm{ \tilde{z}_{t,j}})$ is also equal to $-1$. The choice of $-1$ is consistent with convention in the literature \cite{hoff2002latent,sarkar2005dynamic,raftery2012fast,sweet2020latent} to serve as a baseline for interpretation.

We also mention here that some models add parameter(s) in front of the similarity function in order to link certain effects to the latent space, e.g., in \cite{sewell2015latent}. Such parameters are set to describe the damping/accentuating effect of the distance between two nodes. Exploration of such additional features in our model is beyond the scope of this work and is left for future work.

\section{Simulation Results for Time-invariant Parameters}\label{sec:2}
We evaluate the performance of our proposed posterior-based inference approach through simulation. We consider two sets of parameter settings, one for flocking and the other for polarization (similarly to Figure 4 in the main text), and for each setting we simulate $10$ data sets with number of actors $N=100$ and the number of time points $T=10$. 
Table 1
displays the posterior-based mean and standard deviation for parameter estimation under the two settings. We can see that the parameter estimates are reasonably accurate compared with the truth. 

In our simulation study, we set $\gamma_1^w, \gamma_2^w, \gamma^b \in[-1, 1]$, since in our simulation experience we found that setting their magnitudes too large (typically larger than 1) could lead to divergent model behaviors: we lose the attraction effect since nodes are being pulled too far away from the local averages, although the repulsion effect seems to be well-behaved for large magnitude of $\gamma^b$.

\begin{table}
\caption{Posterior-based mean (standard deviation) for parameters in the flocking (top) and polarization (bottom) settings, based on $N=100$ nodes and $T=10$ time points. [Based on 20 Monte Carlo trials]}
\centering
\tiny
\begin{tabular}{lrrrrrrr}
 & $\hat \alpha$ & $\hat \delta$ & $\hat \gamma_1^w$ & $\hat \gamma_2^w$ & $\hat \gamma^b$ & $\hat \tau$ & AUC \\ \hline
Truth          & 1        & 2        & 0.3    & 0.2 & 0.5    &   1  \\ 
Posterior Mean & 1.122 (0.022)& 2.046 (0.032)& 0.366 (0.085)& 0.273 (0.115)&0.496 (0.108)& 1.058 (0.035)  & 0.830 (0.002) \\
\hline
Truth          & 1        & 3        & 0.7     & 0.2 & -0.5    &   1  \\ 
Posterior Mean & 1.083 (0.083) & 3.055 (0.047) &  0.777 (0.064)&0.205 (0.067)&-0.491 (0.035)&1.039 (0.068)& 0.945 (0.002) \\
\hline
\end{tabular}
\label{tab-estimation}
\end{table}


In the above simulation study, we fix $\sigma^2=1$. To determine how sensitive the estimation procedures are to the true values of $\sigma^2$, we simulate network times series with $N=100$ nodes, $T=10$ time points, under different values of $\sigma \in \{0.01, 0.05, 0.1, 0.5, 0.8, 1, 1.2, 1.5\}$, and estimate the model parameters with $\sigma^2$ fixed at $1$. Such range of $\sigma$ is chosen based on our simulation experience that the nodes would not be able to mimic the desired social dynamics (flocking/polarization) when the noise level is set to be too high, in which case, the nodes movement will be driven predominantly by the noise in the Gaussian autoregressive process in the latent space, not the attractors. This is also consistent with the literature on latent space models of similar type with Gaussian AR process \citep{sewell2015latent,sarkar2005dynamic}, where small values of noise are assumed. 

Table 2
shows the results for the polarization setting. We can see that the estimates for $\gamma^b$ are quite accurate and not sensitive at all to the values of $\sigma$, while all the other parameters are, to varying extents, more sensitive to the values of $\sigma$ although the overall accuracy of the model as modeled through the AUC criterion is always quite high. We note, however, that the signs and relative magnitude are correctly estimated in all cases.

\begin{table}
\caption{Sensitivity Study: under an array of values for $\sigma$, posterior-based mean (standard deviation) for parameters in the polarization setting, based on $N=100$ nodes and $T=10$ time points. [Based on 20 Monte Carlo trials] }
\centering
\scriptsize
\begin{tabular}{lrrrrrrr}
$\sigma$ & $\hat \alpha$ & $\hat \delta$ & $\hat \gamma_1^w$ & $\hat \gamma_2^w$ & $\hat \gamma^b$ & $\hat \tau$ & AUC \\\hline
0.01       &    
1.504 (0.038) &   3.376 (0.034)&   1.065 (0.055)&   0.362 (0.036)&  -0.504 (0.016)&  1.263 (0.054) &  0.960 (0.003)  \\
0.05       &    
1.494 (0.047)&   3.365 (0.043)&  1.073 (0.057)&  0.370 (0.057) & -0.505 (0.018)& 1.258 (0.061)  & 0.960 (0.002)   \\
0.1       &    
1.484 (0.059)&  3.360 (0.040) &  1.064 (0.051)&   0.364 (0.039)&  -0.501 (0.017)&  1.252 (0.051)& 0.960 (0.002)  \\
0.5       &    
1.344 (0.061)&  3.243 (0.043)&  0.926 (0.048)&  0.315 (0.059)&  -0.496 (0.014)&  1.177 (0.054)&  0.955 (0.002)  \\
0.8       &    
1.195 (0.070) &  3.127 (0.050)&  0.814 (0.050)&  0.264 (0.083)&  -0.497 (0.031)&  1.099 (0.054) &   0.948 (0.002)
  \\
1       &    
 1.083 (0.083) &  3.055 (0.047) & 0.777 (0.064)&  0.205 (0.067) & -0.491 (0.035) &  1.039 (0.068)&  0.945 (0.002)
  \\
1.2       &    
0.970 (0.099) & 2.970 (0.050) & 0.762 (0.082) &  0.179 (0.090) & -0.494 (0.035) & 1.006 (0.072) & 0.945 (0.002)
  \\
1.5       &    
 0.843 (0.094)&  2.848 (0.064)&  0.678 (0.099)&  0.100 (0.115)& -0.515 (0.045)&  0.939 (0.051)&  0.947 (0.002)
  \\
  \hline
Truth & 1 & 3 & 0.7 & 0.2  & -0.5 & 1  & \\\hline
\end{tabular}
\label{tab-sensitivity}
\end{table}

\section{Simulation Results for Time-varying Parameters with Change-point}\label{sec:3sim-dic}

In this section, we conduct simulations to evaluate the empirical performance of the proposed parameter estimation procedure for CLSNA models under the time-varying parameter setting with a change-point. In general, we fit two models with time-invariant parameters to the two subperiods of network time series divided by the change-point, respectively, thus obtaining a set of parameter values up to the given change-point, and another set of parameter values after the change-point. The caveat is that the two models cannot be fitted simultaneously since the fitting for the second period should be conditional on the latent positions of the last network in the first fitted model. 

For each trial, we simulate a network time series $y_{1:T}$ with $N=100$ nodes, and length $T=10$, using the following parameter setting. The parameters are set to change at time $6$. In each period: $[1, 5]$, and $[6, 10]$, the parameters are constant respectively. In the first period, $\gamma_1^w$, $\gamma_2^w$ and $\gamma^b $ are set to be $0.6$, $0.6$ and $-0.2$, respectively. In the second period, $\gamma_1^w$, $\gamma_2^w$ and $\gamma^b $ are set to be $0.8$, $0.2$ and $-0.5$, respectively. $\alpha, \delta, \tau$ are set to be fixed all the time.

For each simulated network time series, we fit CLSNA models with constant parameters for the two periods, $[1, r-1]$, and $[r, 10]$, respectively. Since the true change point $k$ is unknown, we fit models for different values of $r$ varying from 4 to 9. For each specified change point $r$, we repeat the trial 20 times and report mean, standard deviation, and DIC based on the 20 replications. The results for parameter estimation are shown in the Table 3.
We can see that for model fitting with the change-point specified at the truth (i.e., r = 6, the bold part), we can get quite good parameter estimates for each period. And such true model has lower DIC value as expected, illustrating that DIC is a useful criteria for model selection when the true change point is unknown.

\begin{table}
\caption{Sub-window analysis with a change-point: posterior-based mean (standard deviation) for parameters estimates, AUC and DIC with specified change-point varying from $4$ to $9$; the true change-point is set at time $6$. The fitting of second subwindow is conditional on the estimated latent positions from the first subwindow. Networks are set to have $N=100$ nodes, and the results are based on 20 Monte Carlo trials.}
\centering
\resizebox{\columnwidth}{!}{%
\begin{tabular}{l|llllllll}\hline
               & $\hat \alpha$ & $\hat \delta$ & $\hat \gamma_1^w$ & $\hat \gamma_2^w$ & $\hat \gamma^b$ & $\hat \tau$ & AUC& DIC  \\ 
 \diagbox{Period}{Truth}           & 1       & 3       & 0.6$\rightarrow$0.8     & 0.6$\rightarrow$0.2 & -0.2$\rightarrow$-0.5    &   1 & \\ \hline
1-3 & 1.137 (0.041)& 3.085 (0.046)& 0.577 (0.175)& 0.608 (0.183)& -0.146 (0.147)& 1.065 (0.058)& 0.856 (0.003) & 35115.168  (783.136)\\
4-10 & 1.086 (0.085)& 3.055 (0.045)& 0.852 (0.056)& 0.422 (0.078)& -0.461 (0.031)&  & 0.945 (0.003)&  \\\hline
1-4 & 1.135 (0.039)& 3.062 (0.034)& 0.554 (0.147)& 0.593 (0.160)& -0.129 (0.139)& 1.064 (0.060)& 0.870 (0.003) & 35092.754  (770.129)\\
5-10 &1.131 (0.056)& 3.076 (0.053)& 0.848 (0.059)& 0.303 (0.041)& -0.475 (0.034)& & 0.950 (0.003)&  \\\hline
\textbf{1-5} & 1.128 (0.043)& 3.060 (0.028)& 0.551 (0.089)& 0.580 (0.117)& -0.123 (0.086)& 1.061 (0.059)& 0.878 (0.003)& \textbf{35093.386} (766.977)\\
\textbf{6-10} &1.113 (0.063)& 3.073 (0.060)& 0.851 (0.056)& 0.209 (0.046)& -0.495 (0.029)&& 0.956 (0.003)&  \\\hline
1-6 & 1.115 (0.040)& 3.058 (0.031)& 0.666 (0.080)& 0.545 (0.102)& -0.251 (0.078)& 1.053 (0.058)& 0.885 (0.003)& 35101.558 (764.716) \\
7-10 &1.122 (0.074)& 3.062 (0.073)& 0.842 (0.060)& 0.195 (0.049)& -0.502 (0.027)& & 0.963 (0.004)& \\\hline
1-7 & 1.108 (0.039)& 3.054 (0.033)& 0.781 (0.072)& 0.539 (0.080)& -0.388 (0.065)& 1.047 (0.058)& 0.895 (0.002)&  35101.899 (770.538) \\
8-10 & 1.124 (0.077)& 3.078 (0.091)& 0.864 (0.078)& 0.180 (0.060)& -0.509 (0.032)& & 0.970 (0.004)&   \\\hline
1-8 & 1.105 (0.037)& 3.057 (0.030)& 0.837 (0.047)& 0.529 (0.061)& -0.452 (0.046)& 1.044 (0.058)& 0.905 (0.003)& 35108.659 (768.719)\\
9-10 & 1.122 (0.105)& 3.067 (0.088)& 0.866 (0.096)& 0.147 (0.101)& -0.505 (0.037)& & 0.974 (0.003)& \\
\hline
\end{tabular}
}
\label{tab-sub-window}
\end{table}

\section{Figures for Edge Persistence}\label{sec:4}
\begin{figure}[H]
\centering
\includegraphics[width=0.5\linewidth]{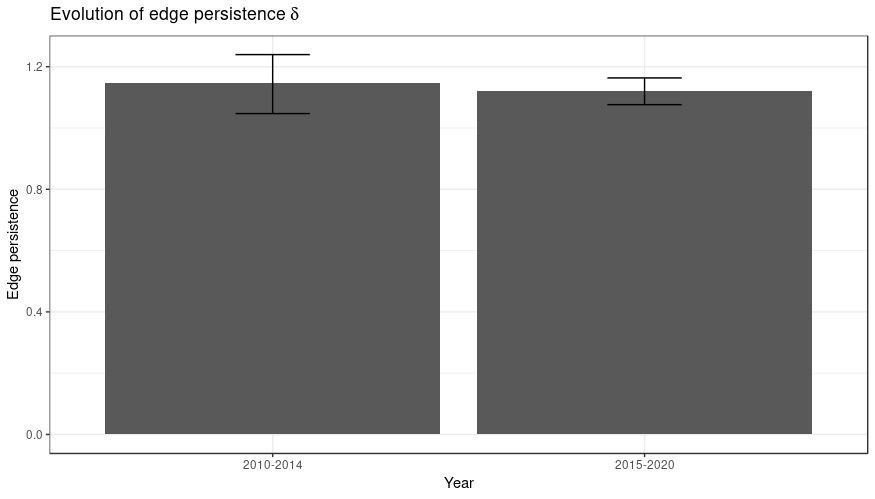}
\caption{Evolution of posterior means and 95\% CI for edge persistence $\delta$ in Twitter congressional hashtag networks, with change-point at year 2015.}
\label{fig_twitter_delta}
\end{figure}

\begin{figure}[H]
\centering
\includegraphics[width=0.5\linewidth]{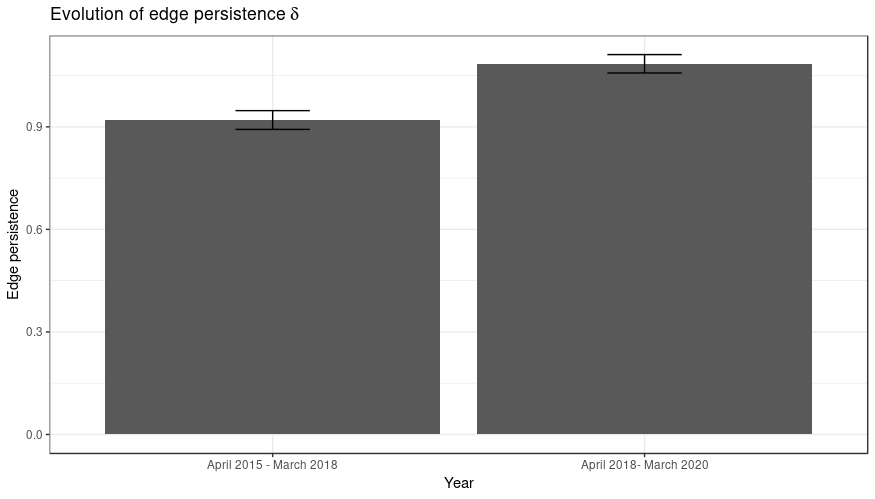}
\caption{Evolution of posterior means and 95\% CI for edge persistence $\delta$ in Reddit comment networks, with change-point at year 2018.}
\label{fig_reddit_delta}
\end{figure}

This section provides two figures mentioned in the main text, for the evolution of edge persistence in Twitter and Reddit data, which are shown in Figure \ref{fig_twitter_delta} and Figure \ref{fig_reddit_delta}. 
 
\section{Supplementary Results in Twitter/Reddit Data Analysis}\label{sec:5}
In this section, we provide the DIC values for all of the competing models in the analysis of Twitter and Reddit networks when choosing change-points using DIC criteria. 
We also present a select handful of alternative analysis results for both networks.
Notice that the change-point preferred by DIC for Twitter data is at 2015, which is a bit earlier than the actual election or presidency year of Trump. 
An alternative hypothesis (a priori) for possible change-point would be 2017, which is the first year of Trump's presidency. 
Therefore, we also present the analysis results with change-point set on 2017, and leave the interpretation to readers. 

\subsection{DIC values}
For twitter data, DIC values for a single change-point chosen at 2012, 2013,..., 2019 are shown in 
Table 4. The lowest DIC value is from model with change-point at 2015.

\begin{table}
\caption{DIC values for competing models with different change-point for Twitter data.}
\centering
\resizebox{\columnwidth}{!}{%
\begin{tabular}{lllllllllll}
& 2012 & 2013 & 2014 & \textbf{2015} & 2016 &  2017 & 2018 &  2019 \\\hline
DIC & 52168.45 &  51186.39 & 50574.23 &  \textbf{50538.64} & 50671.02 & 50727.15 & 50977.68 &  50836.72
\\\hline
\end{tabular}
}
\label{dic-twitter}
\end{table}
For Reddit data, the DIC values for a single change-point chosen at 2017, 2018, 2019 are 254383.18, 253878.32 and 254272.54, respectively. The lowest DIC value is from model with change-points at 2018. 

\subsection{Alternative results for Twitter data with one change-point at 2017}

\begin{figure}[H]
\centering
\includegraphics[width=0.4\linewidth]{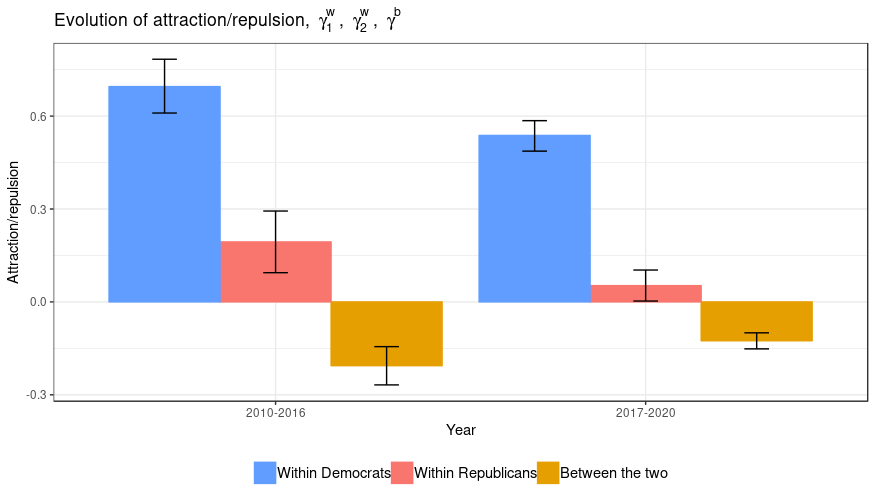}
\includegraphics[width=0.4\linewidth]{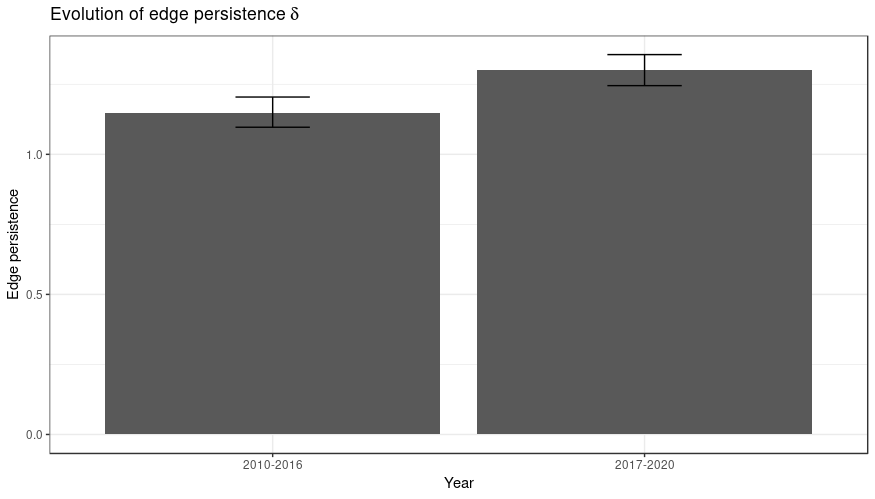}
\caption{Alternative results for Twitter data with one change-point at 2017: evolution of posterior means and 95\% CI for within-group attraction/repulsion $\gamma_1^w$ for Democrats, $\gamma_2^w$ for Republicans, between-group attraction/repulsion $\gamma^b$ and edge persistence $\delta$ in Twitter congressional hashtag networks. In the first period, strong in-party attachment (positive blue, red bar) for both parties, between-group repulsion (yellow bar) appeared but did not dominate. In the second period, within-group attraction decreased for both parties; the out-party negative feelings dominated the in-party positive feelings for Republicans (magnitude of yellow greater than that of red); and the between-group polarization decreased in the second time period.}
\label{fig_twitter_2017}
\end{figure}

Although the model with one change-point at 2015 was selected based on DIC values, it is still interesting to see the results with change-point set on 2017 - the first year under Trump's administration. The results are shown in Figure \ref{fig_twitter_2017}, which do not tell a much different story from the model with change-point at 2015: the positive feelings about their own for both parties fell in the second time period, and for Republican members of Congress, negative feelings about the other party started to dominate positive feelings about their own. What's intriguing is the decreased trend of polarization under the Trump administration years, as opposed to the indication of increased trend from change-point on 2015. 

\subsection{Alternative results for Reddit data with one change-points at 2017}
\begin{figure}[H]
\centering
\includegraphics[width=0.4\linewidth]{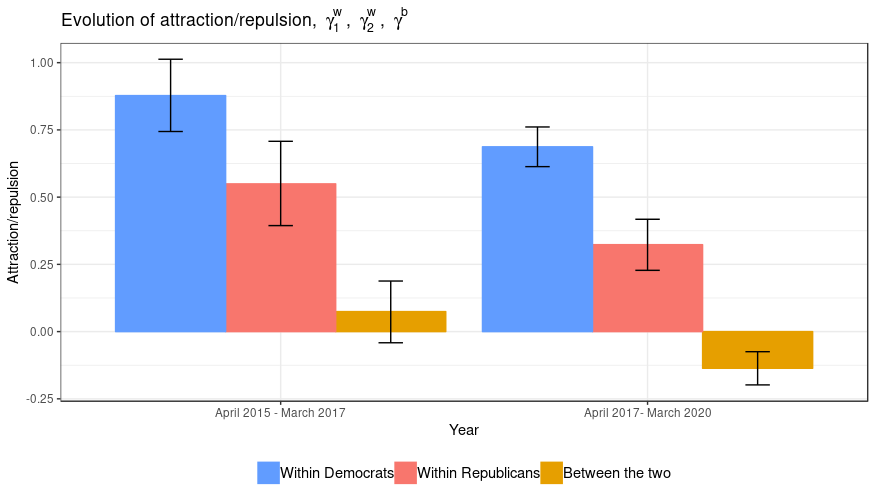}
\includegraphics[width=0.4\linewidth]{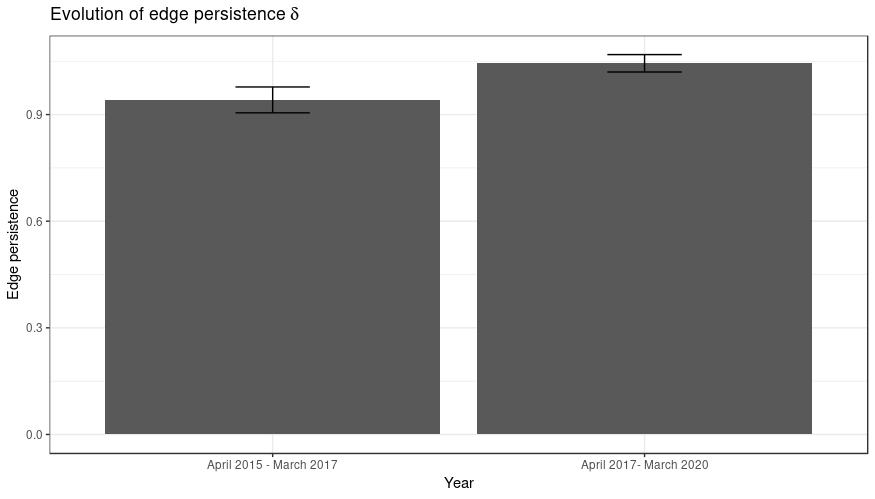}
\caption{Alternative results for Reddit data with one change-points at 2017: evolution of posterior means and 95\% CI for within-group attraction $\gamma_1^w$ for Democrats, $\gamma_2^w$ for Republicans, between-group attraction/repulsion $\gamma^b$, and edge persistence $\delta$ in Reddit comment networks. In this first period, strong in-party attachment for both parties, no evidence of between-group polarization. In the second period, within-group attraction decreased for both parties, between-group repulsion (yellow) appeared but did not dominate.}
\label{fig_reddit_2017}
\end{figure}

The results for Reddit data with one change-point at 2017 are shown in Figure \ref{fig_reddit_2017}, which are qualitatively similar to that with change-point in 2015, except the indication of trend for polarization. 

\subsection{Alternative results for Twitter data with two change-points at 2014 and 2019} 
\begin{figure}[H]
\centering
\includegraphics[width=0.4\linewidth]{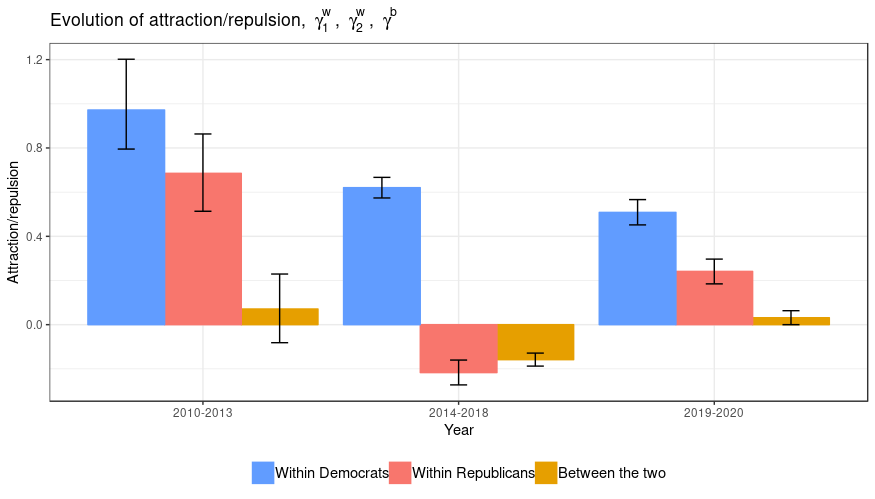}
\includegraphics[width=0.4\linewidth]{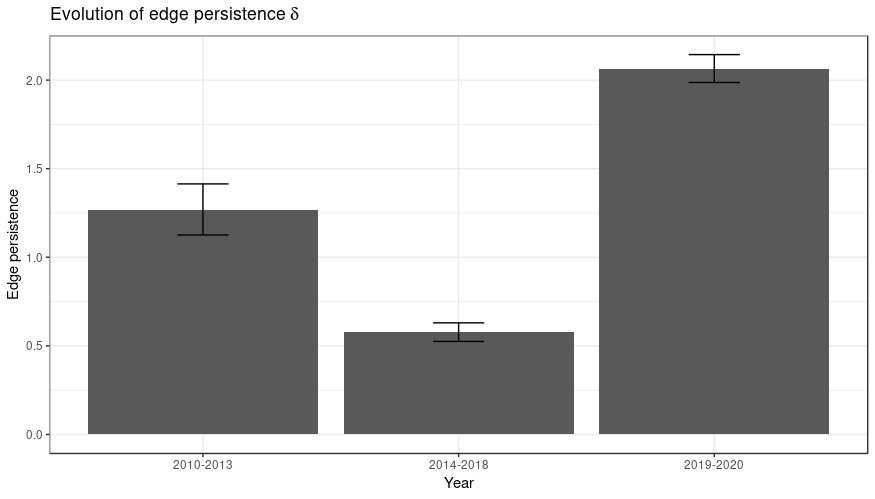}
\caption{Alternative results for Twitter data with two change-points at 2014 and 2019: evolution of posterior means and 95\% CI for within-group attraction/repulsion $\gamma_1^w$ for Democrats, $\gamma_2^w$ for Republicans, between-group attraction/repulsion $\gamma^b$ and edge persistence $\delta$ in Twitter congressional hashtag networks. In the first period, strong in-party attachment for both parties (positive blue red), less evidence for the existence of polarization. In the second period, both between-group repulsion (negative yellow), and within-group repulsion for Republicans (negative red) appear. In-party negative feelings dominates out-party negative feelings (magnitude of red greater than that of yellow). For the third period, both in-/out-party negative feeling disappear.}
\label{fig_cp2_twitter}
\end{figure}

We also present results for Twitter data with two change-points at 2014 and 2019 in Figure \ref{fig_cp2_twitter}, which is chosen with DIC criteria among ${8 \choose 2} = 28$ competing models that place two change-points between 2012 and 2019. Notice that the DIC values for the selected models with zero-, one-, and two-change-point are 53221, 50538 and 50249, respectively, which suggests a much smaller improvement from one change-point to two change-points than that from zero change-point to one change-point.
Given that the data is only at yearly resolution, it seems more practical to focus on the model with one change-point. However, we present the results here to demonstrate the possibility of a richer story that could be told potentially from models with more change-points, which of course needs deeper analysis with networks at finer resolution.

\section{On the robustness of model results to network
construction}\label{sec:6}

To explore the robustness of the model results to the network construction, we fit models on both Twitter and Reddit networks that are constructed under different edge binarization strategies and compare the results. 

\begin{figure}
\centering
\includegraphics[scale=0.3]{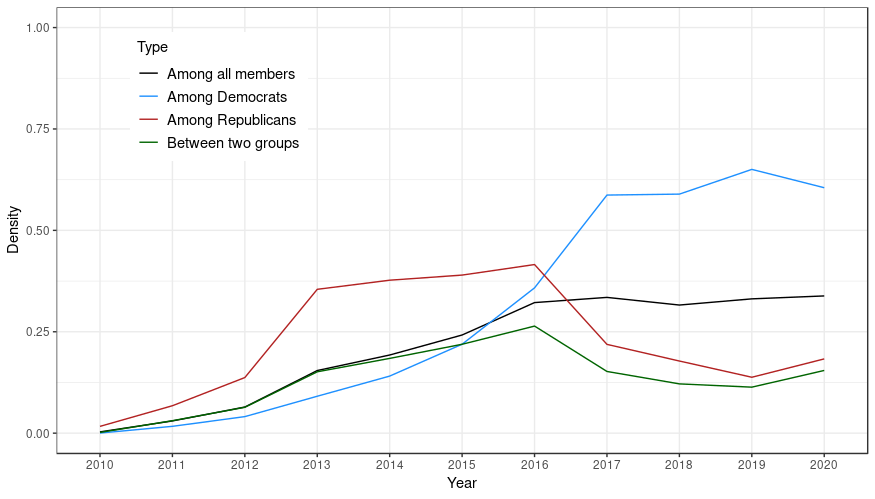}
\includegraphics[scale=0.3]{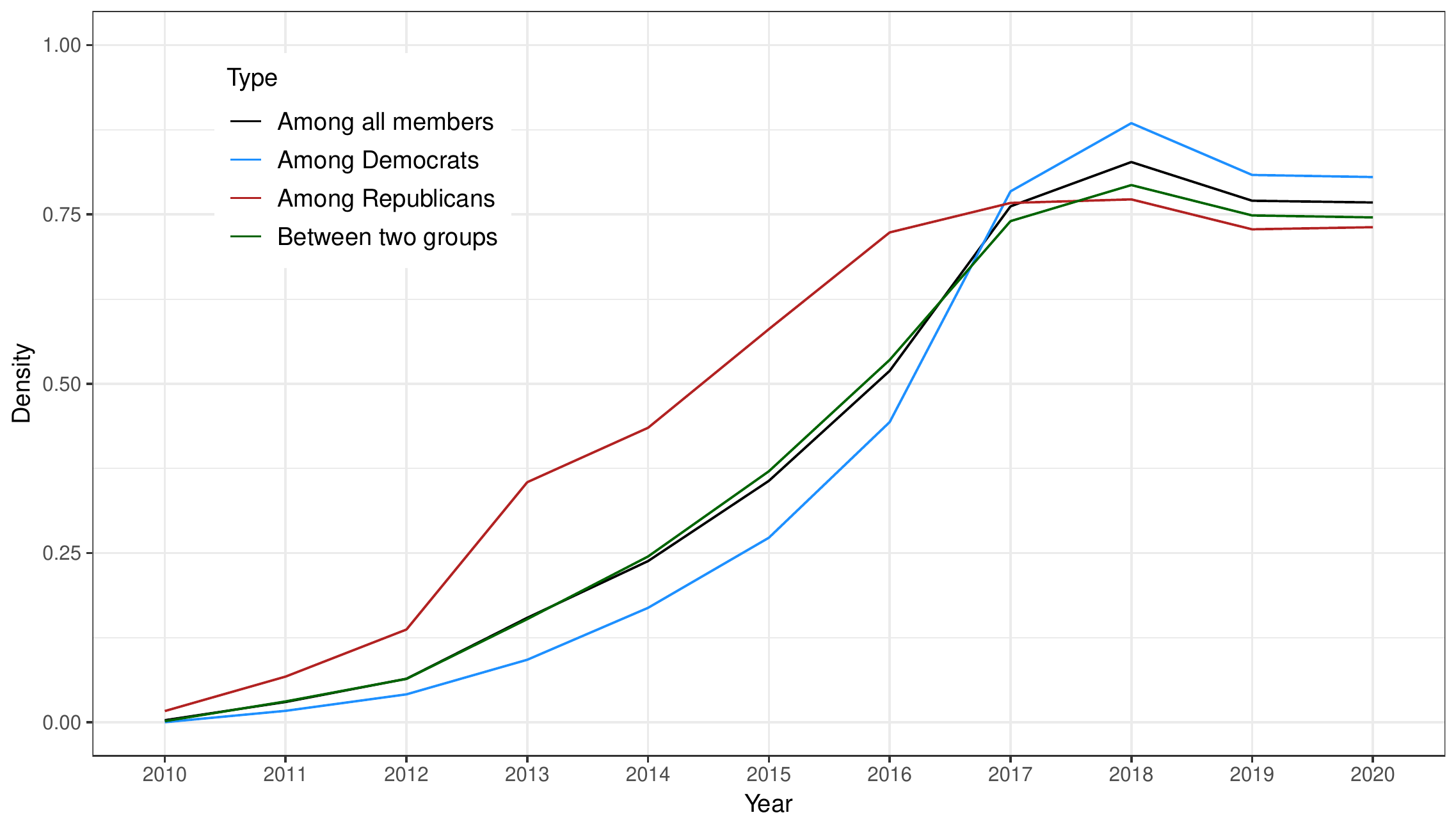}
\includegraphics[scale=0.3]{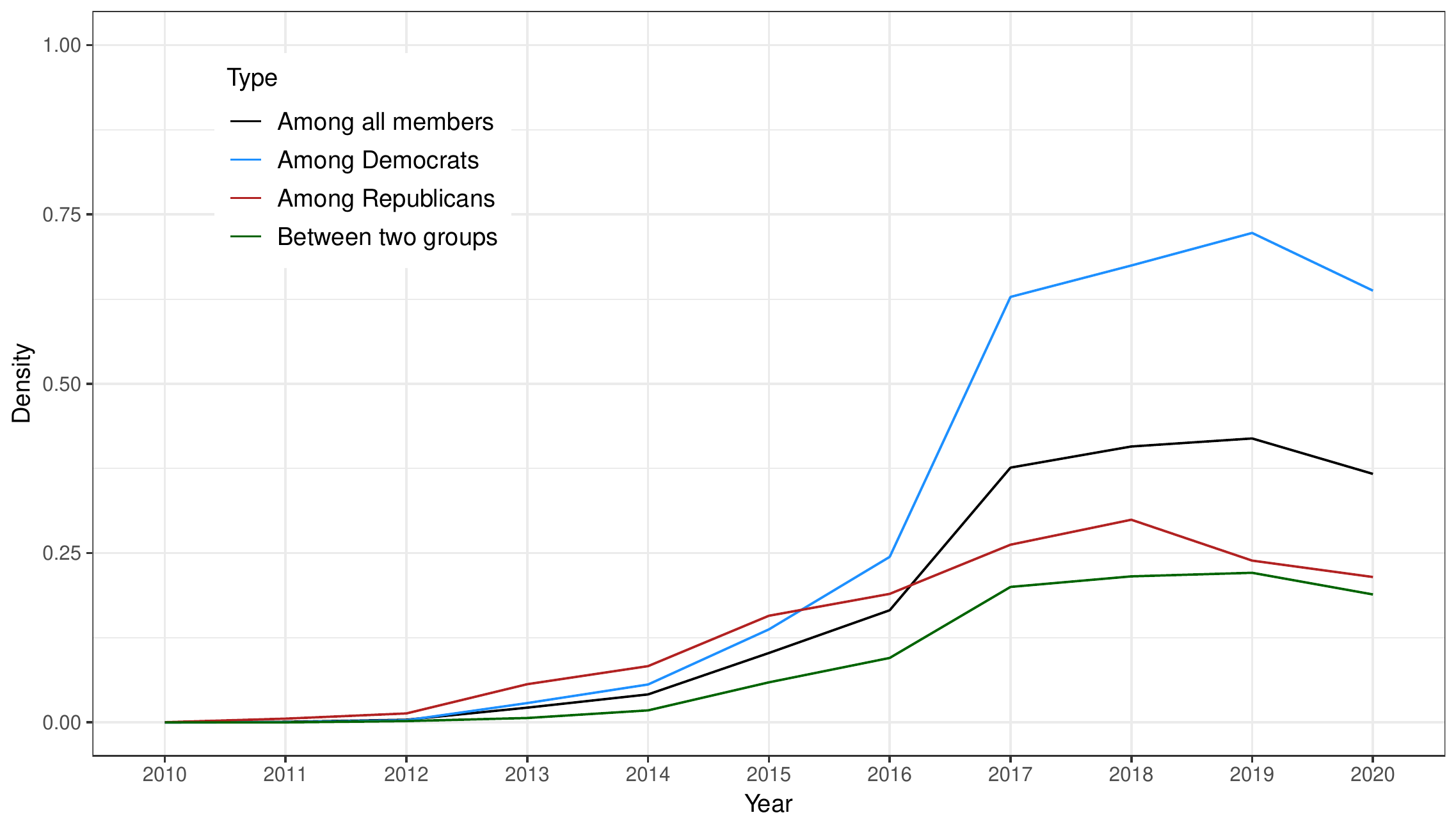}
\caption{Densities over years for constructed Twitter networks. \textit{Upper left}: networks are constructed with dynamic mean threshold (the explicit thresholds used for each year are 0.01, 0.10, 0.26, 0.88, 1.49, 3.24, 4.97, 12.46, 14.28, 14.95, 11.24, which is the average number of common hashtags tweeted by members of congress each year). \textit{Upper right}: network are constructed with static threshold value: 0. \textit{Bottom}: network are constructed with static threshold value: 10.}
\label{fig1}
\end{figure}

For Twitter data, the network sequence analyzed in the main text are constructed using a dynamic thresholding strategy, i.e., we connect the two congress members with an edge in a year if the common hashtags they tweeted were more than the average number of common
hashtags tweeted by any two members of congress that year; otherwise we leave them unconnected. Alternatively, we construct networks using a static thresholding strategy, where two congress members are connected with an edge if the common hashtags they tweeted were more than a fixed threshold value. We construct two new sequences of networks under threshold value 0 and 10. Figure \ref{fig1} shows a comparison on densities of networks constructed using the three different strategies. The model fitting results for the two newly constructed networks are shown in Table 6 and Table 7.

\begin{figure}[H]
\centering
\includegraphics[scale=0.3]{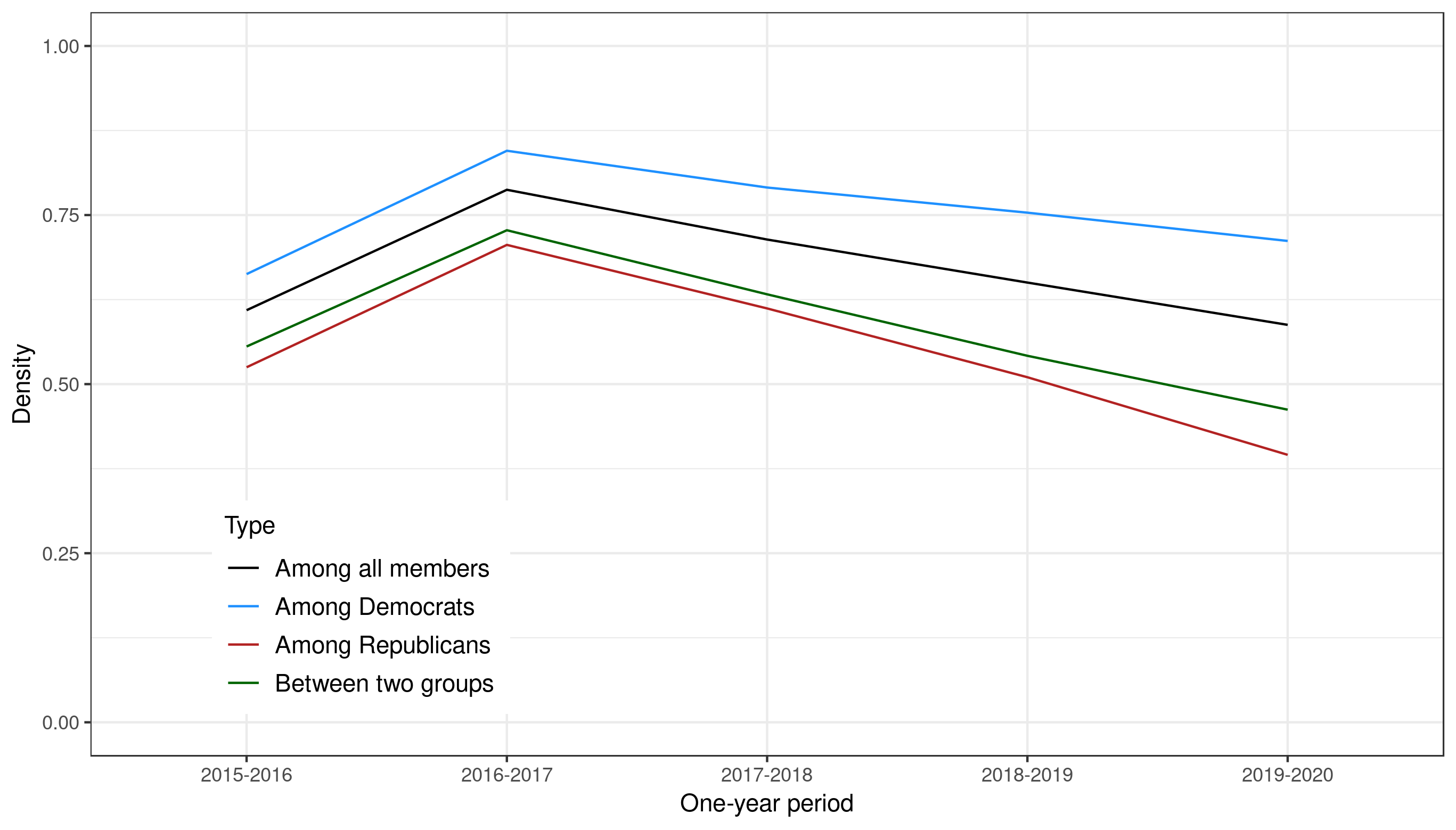}
\includegraphics[scale=0.3]{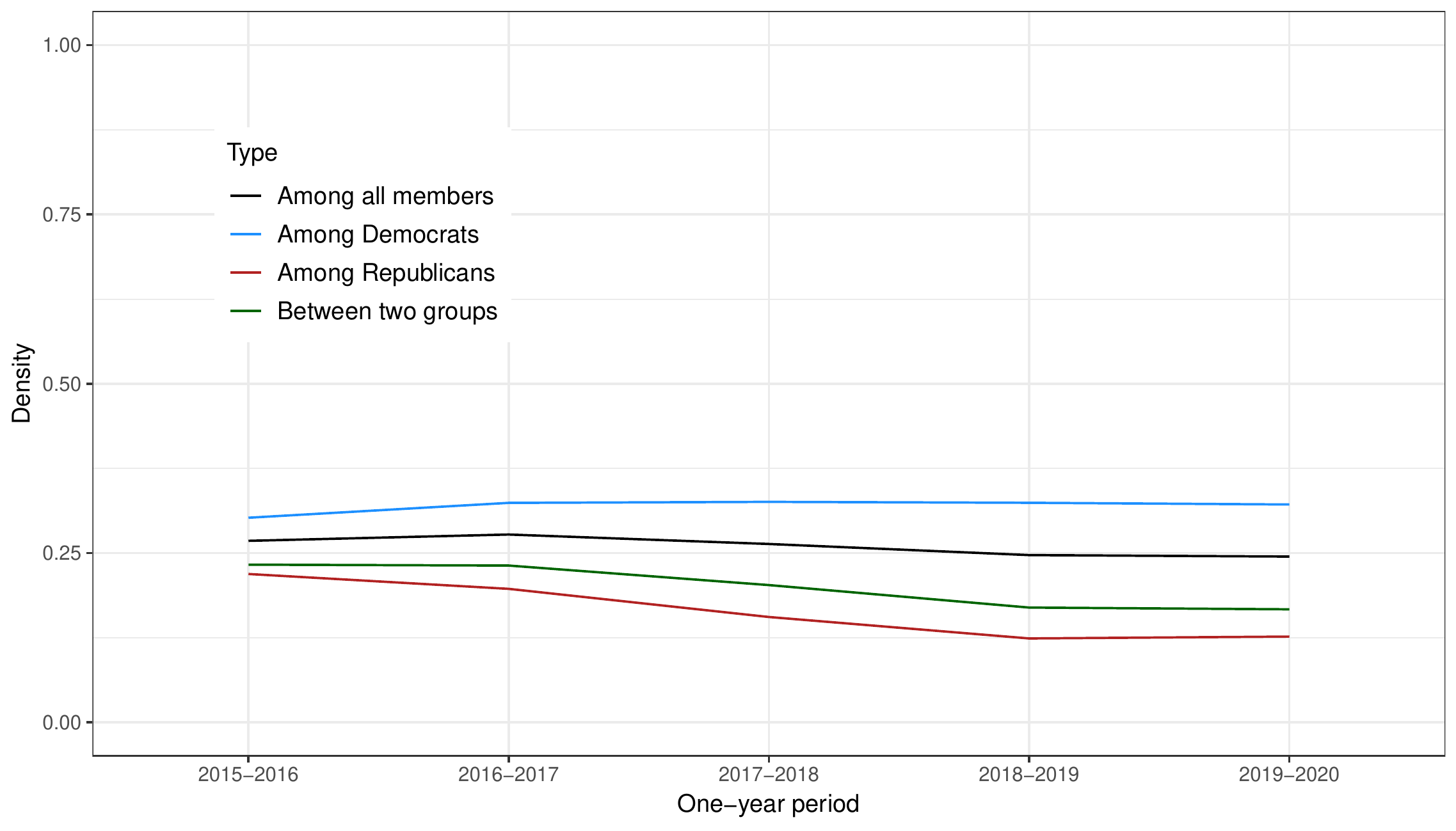}
\caption{Densities over years for constructed Reddit networks. \textit{Left}: networks are constructed with static zero threshold. \textit{Right}: network are constructed with dynamic mean threshold (the explicit thresholds used for each year are: 2.35, 6.03, 5.44, 4.47, 3.19, which is the average number of posts that two users both commented on each year).}
\label{fig2}
\end{figure}

For Reddit data, the network sequence analyzed in the paper are constructed using a static thresholding strategy, i.e., we connect two users with an edge in a year if they both commented on at least one post that year. Alternatively, we construct networks using a dynamic thresholding strategy, where two users are connected with an edge in a year if the number of posts they both commented on was more than the average that year. The network density plot and the corresponding results for the networks constructed this way are shown in Figure \ref{fig2} (right) and Table 8. 

Table 5
below gives an summary on whether each qualitative conclusions in the main texts holds under different network constructions. 

\begin{table}
\caption{Summary of impact of network constructions on qualitative results}
\label{tab0}
\resizebox{\columnwidth}{!}{%
\begin{threeparttable}
\begin{tabular}{l|lll|ll}
& \multicolumn{3}{c|}{Twitter}                            & \multicolumn{2}{c}{Reddit}        \\ \hline
\multicolumn{1}{c|}{Conclusion} & dynamic thre       & static thre: 0 & static thre: 10 & static thre: 0    & dynamic thre \\
 & (results in paper) & & & (results in paper) &  \\
A &  \checkmark & \checkmark     &    \checkmark  &   \checkmark   & \checkmark \\
B &  \checkmark  & \checkmark &  \checkmark    &\checkmark     &  \checkmark \\
C&   \checkmark   & \xmark   & \checkmark  &  \checkmark   &   \xmark \\
D & \checkmark   & \checkmark   &  \checkmark   & \checkmark    &   \checkmark\\
E & \checkmark    &  \xmark    &   \xmark   & \xmark      & \xmark \\\hline
\end{tabular}
\begin{tablenotes}
\item Conclusions in the main text:
    \item A: presence of edge persistence ($\hat \delta > 0$)
    \item B:  higher within-group attraction for Democrats than Republicans ($\hat \gamma_1^w > \hat \gamma_2^w$)
    \item C:  presence of between-group repulsion ($\hat \gamma^b <0 $)
    \item D: smaller magnitude of between-group repulsion than within-group attraction for Democrats ($|\hat \gamma^b| < |\hat \gamma_1^w|$)
    \item E: greater magnitude of between-group repulsion than within-group attraction for Republicans ($|\hat \gamma^b| > |\hat \gamma_2^w|$)
\end{tablenotes}
\end{threeparttable}
}
\end{table}

For the Twitter data, the qualitative conclusions A, B and D are quite robust to the network construction. The conclusions on the presence of polarization and on the dominance of between-group repulsion for Republicans seem to be sensitive to the network construction. However, given that the hashtag usage in Twitter by members of Congress increased substantially over the past decade, and the fact that the average number of common hashtags tweeted increased from $0.01$ in 2010 and $11.24$ in 2020, we think it is more appropriate to use the yearly average in defining edges instead of a static threshold. This choice allows our model to capture more meaningful interaction dynamics beyond increasing hashtag usage. 

For the Reddit data, all qualitative conclusions are robust except the one on the presence of polarization. In contrast to the Twitter case, average commenting rates in Reddit did not change substantially over time. Also as shown in Figure \ref{fig2}, the networks constructed with threshold zero exhibits more varying interaction dynamics than the other one. Consequently,  it is more appropriate to use a static threshold in defining edges for the Reddit data. 

\begin{table}
\caption{Static threshold: 0. Summary statistics for the posterior distribution of parameters using the whole sequence of Twitter networks from 2010 to 2020}
\label{twitter-tab2}
\begin{threeparttable}
\centering
\begin{tabular}{llllll}
 & $\hat \alpha$ & $\hat \delta$ & $\hat \gamma_1^w$ & $\hat \gamma_2^w$ & $\hat \gamma^b$\\\hline
Mean & 3.556 & 1.731 & 0.585 & 0.542 & 0.011 \\
SD & 0.024 & 0.024 & 0.038 & 0.035 & 0.024 \\ 
2.5\% Quantile &3.511 & 1.683 & 0.510 & 0.473  & -0.036 \\ 
97.5\% Quantile & 3.605 &1.778  &0.660 & 0.611 & 0.057  \\ 
\hline
\end{tabular}
\begin{tablenotes}
\item 1-Democrats, 2-Republicans. These results indicate presence of edge persistence ($\hat \delta > 0$), higher within-group attraction for Democrats than Republicans ($\hat \gamma_1^w > \hat \gamma_2^w$ (mean difference = .043, SE = .046, P(difference$>$0)=0.828)), no evidence of between-group repulsion ($P(\hat \gamma^b <0) = 0.321$)), and smaller magnitude of between-group force than within-group attraction for both Republicans and Democrats ($|\hat \gamma^b| < |\hat \gamma_1^w|$ (mean difference = -.565, SE = .045), $|\hat \gamma^b| < |\hat \gamma_2^w|$ (mean difference = -.521, SE = .043).
\end{tablenotes}
\end{threeparttable}
\end{table}

\begin{table}
\caption{Static threshold: 10. Summary statistics for the posterior distribution of parameters using the whole sequence of Twitter networks from 2010 to 2020}
\label{twitter-tab3}
\begin{threeparttable}
\centering
\begin{tabular}{llllll}
 & $\hat \alpha$ & $\hat \delta$ & $\hat \gamma_1^w$ & $\hat \gamma_2^w$ & $\hat \gamma^b$\\\hline
Mean & 2.472 & 1.969 & 0.594 & 0.262 & -0.066 \\ 
SD &  0.020 & 0.021 & 0.022 & 0.023 & 0.014 \\ 
2.5\% Quantile & 2.433 & 1.928 & 0.551& 0.215 & -0.094\\ 
97.5\% Quantile & 2.512 & 2.012 & 0.638 & 0.307  & -0.038 \\ 
\hline
\end{tabular}
\begin{tablenotes}
\item 1-Democrats, 2-Republicans. These results indicate presence of edge persistence ($\hat \delta > 0$), higher within-group attraction for Democrats than Republicans ($\hat \gamma_1^w > \hat \gamma_2^w$ (mean difference = .333, SE = .033, P(difference$>$0)=1)), presence of between-group repulsion ($P(\hat \gamma^b <0)=1 $), and smaller magnitude of between-group repulsion than within-group attraction for both Republicans and Democrats ($|\hat \gamma^b| < |\hat \gamma_1^w|$ (mean difference = -.529, SE = .026), $|\hat \gamma^b| < |\hat \gamma_2^w|$ (mean difference = -.196, SE = .023). 
\end{tablenotes}
\end{threeparttable}
\end{table}

\begin{table}
\caption{Dynamic thresholding with mean. Summary statistics for the posterior distribution of parameters using the whole sequence of Reddit networks from 2015 to 2020}
\begin{threeparttable}
\centering
\begin{tabular}{llllll}
& $\hat \alpha$ & $\hat \delta$ & $\hat \gamma_1^w$ & $\hat \gamma_2^w$ & $\hat \gamma^b$  \\ 
\hline
Mean & 1.275 & 1.400 & 0.302  & 0.151 & 0.007 \\ 
  SD & 0.010 & 0.011 & 0.020 & 0.034 & 0.021 \\ 
2.5\% Quantile & 1.255 & 1.379  &  0.264 & 0.084 & -0.034 \\ 
97.5\% Quantile & 1.293 &1.421   & 0.341 & 0.217 &  0.050 \\ 
\hline
\end{tabular}
\begin{tablenotes}
\item 1-Democrats, 2-Republicans. This results indicate presence of edge persistence ($\hat \delta > 0$), higher within-group attraction for Democrats than Republicans ($\hat \gamma_1^w > \hat \gamma_2^w$ (mean difference = .151, SE = .031, P(difference$>$0)=1)), no evidence of between-group repulsion ($P(\hat \gamma^b <0) = 0.365$)), and less magnitude of between-group force than within-group attraction for both Democratic and Republican Reddit users ($|\hat \gamma^b| < |\hat \gamma_1^w|$ (mean difference = -.284, SE = .027), $|\hat \gamma^b| < |\hat \gamma_2^w|$ (mean difference = -.133, SE = .040).
\end{tablenotes}
\end{threeparttable}
\label{tab5}
\end{table}

\clearpage

\bibliographystyle{rss}
\bibliography{supplement, cantay_bib}